\documentclass[10pt,twocolumn]{article}

\usepackage[T1]{fontenc}
\usepackage{lmodern}                   
\usepackage[margin=2.1cm]{geometry}
\usepackage{amsmath,amssymb}
\usepackage{graphicx}
\usepackage{xcolor}
\usepackage{mathbbol}
\usepackage{booktabs}
\usepackage{algorithm}
\usepackage{algorithmic}
\usepackage{ifthen}
\usepackage{microtype}
\usepackage{xurl}                      
\usepackage{etoolbox}
\usepackage[colorlinks=true,allcolors=blue]{hyperref}

\apptocmd{\thebibliography}{\raggedright}{}{}

\makeatletter
\renewcommand\section{\@startsection{section}{1}{\z@}%
  {-3.5ex \@plus -1ex \@minus -.2ex}{2.3ex \@plus.2ex}%
  {\normalfont\Large\bfseries\raggedright}}
\renewcommand\subsection{\@startsection{subsection}{2}{\z@}%
  {-3.25ex\@plus -1ex \@minus -.2ex}{1.5ex \@plus .2ex}%
  {\normalfont\large\bfseries\raggedright}}
\renewcommand\subsubsection{\@startsection{subsubsection}{3}{\z@}%
  {-3.25ex\@plus -1ex \@minus -.2ex}{1.5ex \@plus .2ex}%
  {\normalfont\normalsize\bfseries\raggedright}}
\makeatother

\newcommand{\appendixstart}{%
  \appendix
  \renewcommand{\thesection}{Appendix~\Alph{section}}%
  \setcounter{figure}{0}%
  \renewcommand{\thefigure}{\Alph{section}\arabic{figure}}%
  \setcounter{table}{0}%
  \renewcommand{\thetable}{\Alph{section}\arabic{table}}%
  \setcounter{equation}{0}%
  \renewcommand{\theequation}{\Alph{section}\arabic{equation}}%
}

\newcommand{\Tr}{\mathrm{Tr}}          

\newcommand{\dv}[2]{\frac{\mathrm{d} #1}{\mathrm{d} #2}}
\newcommand{\comm}[2]{\left[ #1, #2 \right]}

\newcommand{\dyad}[2][]{%
  \ifthenelse{\equal{#1}{}}%
    {\left| #2 \right\rangle \left\langle #2 \right|}%
    {\left| #1 \right\rangle \left\langle #2 \right|}%
}

\title{\vspace{-2em}Interpolation of unitaries with time-dependent Hamiltonians\\ via Deep Learning}
\date{\today}

\begin{document}

\twocolumn[{%
\centering
{\LARGE\bfseries Interpolation of unitaries with time-dependent Hamiltonians via Deep Learning\par}
\vspace{1.2em}
{\large Antonio Guerra,$^{1,*}$ Daniel Uzcategui-Contreras,$^{2}$ Aldo Delgado,$^{1}$ Esteban S. G\'omez$^{1}$\par}
\vspace{0.9em}
\begin{minipage}{0.86\textwidth}\centering\small\itshape
$^{1}$Departamento de F\'isica e Instituto Milenio de Investigaci\'on en \'Optica, Universidad de Concepci\'on, Casilla 160-C, Concepci\'on, Chile\\[0.2em]
$^{2}$Departamento de Matem\'atica y F\'isica Aplicadas, Universidad Cat\'olica de la Sant\'isima Concepci\'on, Alonso de Ribera 2850, Concepci\'on, Chile
\end{minipage}
\vspace{0.9em}
{\small$^{*}$\,\href{mailto:antongonzalez@udec.cl}{antongonzalez@udec.cl}\par}
\vspace{1.6em}
\begin{minipage}{0.88\textwidth}
\noindent\textbf{Abstract.} Quantum systems governed by time-dependent Hamiltonians pose significant challenges for the accurate computation of their unitary time-evolution operators. In this work, we introduce a deep learning approach based on Physics-Informed Neural Networks to estimate these operators over the full time domain, either by predicting the time-evolution operator directly or by predicting an effective Hamiltonian from which that operator is computed. Unitarity is imposed as a constraint in the first case, and holds by construction in the second, where a propagation layer based on the second-order Magnus expansion or Trotterization maps the predicted Hamiltonian to the evolution operator. The model is trained on characterization data from a small number of times, generated numerically here but standing in for measurements on a system whose Hamiltonian is unknown to the model. The loss combines three contributions: agreement with pairs of input and evolved states, agreement with the evolution operator at the sampled times when it is available, and a unitarity constraint that requires no measured data. Systems of 2 to 8 qubits are considered, with a fully non-zero Pauli Hamiltonian up to 6 qubits and nearest-neighbor Ising and $XYZ$ Heisenberg models beyond. Across all system sizes, and using a limited number of sampled times, a statistical analysis over independent Hamiltonian realizations gives mean gate fidelities close to $0.99$ with low variability. Comparable fidelities are obtained when the model is trained from state data alone, without providing the evolution operator at any sampled time during training. Once trained, the model interpolates the evolution operator with inference times $3.7\times$ to $28.8\times$ lower than those of a numerical interpolation baseline based on geodesic interpolation, depending on system size and propagation method. These results support the use of the approach for data-driven simulation and identification of quantum dynamical systems.
\end{minipage}
\vspace{2.2em}
}]

\section{Introduction}
\label{sec:intro}

Over the past decade, advances in Artificial Neural Networks and Deep Learning have enabled solutions to computational problems previously considered intractable~\cite{terven2025, sarker2021}, with growing impact across the physical sciences. Within physics, neural networks have been successfully applied to quantum state reconstruction from measurements~\cite{torlai2018, cha2022}, the combination of marginals into global quantum states~\cite{uzcateguicontreras2024}, and the solution of differential equations through Physics Informed Neural Networks (PINNs)~\cite{raissi2019}. These developments illustrate the versatility of machine learning for modeling and simulating complex quantum systems, motivating their application to the dynamics of time-dependent Hamiltonians studied in this work.

A fundamental challenge in quantum theory is the accurate modeling of the dynamics of quantum systems driven by time-dependent Hamiltonians. Unlike the time-independent case, where the time-evolution operator reduces to a simple matrix exponential $e^{-iHt}$, time-dependent Hamiltonians give rise to the time-ordering operator, which precludes a closed-form solution and introduces substantial mathematical and numerical complexity. The cost of any direct approach is further compounded by the exponential growth of the Hilbert space~\cite{lloyd1996}, together with the rapidly growing number of terms required in the series expansions used to represent the time-dependent evolution operator $U(t)$: for an $N$-qubit system the evolution operator is a $2^N \times 2^N$ matrix, and its dense construction or exponentiation scales as $\mathcal{O}(2^{3N})$. These expansions, such as the Dyson or Magnus series, involve nested time integrals whose number grows with the expansion order, and the associated numerical errors tend to accumulate over long evolution times~\cite{wiebe_simulating_2011, blanes2009}. Even on quantum hardware, simulating time-dependent dynamics is generally more demanding than the time-independent case, particularly for rapidly varying or highly oscillatory Hamiltonians~\cite{an2022}. Despite these difficulties, time-dependent Hamiltonians play a central role in quantum technologies, including quantum control and gate optimization~\cite{Ansel_2024}, as well as quantum simulation of complex many-body systems~\cite{georgescu2014, cao2025}.

Several approaches have been developed to address this challenge. Methods based on truncated Dyson series expansions have shown promising results in handling the time-ordering operator~\cite{berry_2015, krieferova_2019}. Magnus-expansion-based algorithms have proven particularly effective in quantum control~\cite{Dalgaard_2022}, and recent work has extended these to highly-oscillatory Hamiltonians relevant to near-term quantum devices~\cite{chen_quantum_2023, fang_time-dependent_2024}. Modified Suzuki-Trotter decompositions have been proposed for interpolating unitary operators in time-dependent settings~\cite{schilling2024}. Tensor-network methods provide a complementary route: they enable efficient classical simulation of one-dimensional systems through time-evolving block decimation~\cite{vidal_efficient_2004}, and recent constructions optimize Suzuki-Trotter decompositions for Hamiltonian simulation and adiabatic quantum computing~\cite{tepaske2023, causer2024}. On the machine learning side, PINNs have been explored to learn quantum dynamics under physical constraints~\cite{norambuena2024}, as well as to reconstruct time-dependent Hamiltonians using deep learning models~\cite{che_2021, han_2021}. Other measurement-based protocols have also been proposed for this task~\cite{franceschetto2025}.

The setting we address is that of a quantum system with an unknown time-dependent Hamiltonian, accessible only through characterization at a finite set of times $\{t_i\} \subset [0,T]$. From the time-evolution operators $U(t_i)$ available at those times, and/or the associated evolved states $(\rho_0, \rho_{t_i})$, we seek to interpolate $U(t)$ over the entire interval $[0,T]$. The model takes a single scalar, the time $t$, as input and returns the corresponding time-evolution operator, learning the dynamics from the characterization data alone. Throughout this work these data are generated numerically, but they play the role of experimental data obtained, for instance, by quantum process tomography (QPT). Each model is trained to represent the dynamics of a single Hamiltonian realization; learning a mapping that generalizes across different Hamiltonians is beyond the scope of this work.

Our method is based on the PINN framework, in which known physical constraints are explicitly incorporated into the training process to guide the learning of physically consistent solutions. Due to the exponential growth of the Hilbert space with the number of qubits, we train separate models per system size and test two distinct model architectures: (i) a model trained to directly predict the unitary operator $U(t)$, and (ii) a model that predicts the coefficients of an effective time-dependent Hamiltonian $H_\mathrm{eff}(t)$, from which $U(t)$ is subsequently computed via the Magnus expansion or Trotterization. The motivation for the second architecture is scalability: by reducing the task from learning the full unitary to learning an effective Hamiltonian, the number of free parameters is drastically reduced, lowering the computational resources required for training.

To assess generality, for systems of up to 6-qubits we consider a fully non-zero Pauli Hamiltonian, the most general $N$-qubit Hamiltonian. This choice serves as a proof of principle: an architecture expressive enough to represent the dynamics of such a general Hamiltonian is, a fortiori, expressive enough to represent that of any sparser, physically motivated Hamiltonian, although each case requires training a dedicated model on its own data. For systems of 7- and 8-qubits, computational constraints require restricting the analysis to two physically motivated models: a nearest-neighbor Ising Hamiltonian and a nearest-neighbor $XYZ$ Heisenberg Hamiltonian, both of which are directly relevant to experimentally realizable quantum platforms~\cite{Kim2011, Kim2012, Kadowaki1998}.

A notable feature of the proposed framework is its ability to achieve high gate fidelity across the entire time domain, even when trained on datasets with a small number of temporal samples. Models trained with coarse temporal discretizations (e.g., $\Delta t = 1.0$) exhibit fidelities comparable to those trained with finer grids (e.g., $\Delta t = 0.1$), demonstrating strong interpolation capability. To assess robustness, we further perform a statistical analysis in which many independent models are each trained on a different randomly sampled Hamiltonian realization; the resulting models consistently achieve high average fidelity with low variability across realizations. Furthermore, the trained models exhibit lower inference times than conventional numerical methods such as geodesic interpolation~\cite{schilling2024}, suggesting that the proposed approach may be particularly useful in scenarios where the time-evolution operator must be evaluated repeatedly. A representative example is dynamical quantum state tomography~\cite{Siva2022, peruzzo2026}, where a well-characterized dynamical map is a prerequisite of the protocol.

This article is organized as follows. In Section~\ref{sec:teo}, we present the theoretical background. Section~\ref{sec:dl} introduces the deep learning models used in this work, including their architectures, training procedures, and performance metrics. In Section~\ref{sec:results}, we analyze the accuracy of the proposed approach for different system sizes. In Section~\ref{sec:time-benchmark}, we compare the inference time of the models against conventional methods. Section~\ref{sec:conc} summarizes our conclusions and outlines potential future research directions. Finally, \ref{sec:app1} provides a detailed description of the model architecture. All source code used in this work is publicly available on GitHub.\footnote{\url{https://github.com/guerra-antonio/PINN_dynamics_estimation}}

\section{Quantum dynamics with time-dependent Hamiltonians}
\label{sec:teo}

The state of a quantum system is described by a density operator $\rho$ which is a normalized ($\mathrm{Tr}(\rho)=1$) positive semi-definite ($\rho \ge 0$) linear operator. This operator provides a complete description of a quantum system. An isolated quantum system, when subjected to an interaction described by a time-dependent Hamiltonian $H(t)$, evolves according to the von Neumann equation
\begin{equation}
  \label{eq:von-neumann-eq}
  \dv{}{t}\rho(t) = -\frac{i}{\hbar}\ [H(t),\rho(t)],
\end{equation}
where $\hbar$ is the reduced Planck constant. The general solution of Eq.~\eqref{eq:von-neumann-eq} is given by
\begin{equation}
  \label{eq:state-evolution}
  \rho (t) = U(t, t_0)\ \rho(t_0)\ U^{\dagger}(t, t_0),
\end{equation}
where the unitary evolution operator is given by the infinite expansion
\begin{eqnarray}
\label{eq:dyson}
U(t, t_0) &=& \mathbb{1} - \frac{i}{\hbar} \int_{t_0}^{t} dt_1\, H(t_1) \nonumber \\
&& + \left( -\frac{i}{\hbar} \right)^2 \int_{t_0}^{t} dt_1 \int_{t_0}^{t_1} dt_2\, H(t_1) H(t_2) \nonumber \\
&& + \left( -\frac{i}{\hbar} \right)^3 \int_{t_0}^{t} dt_1 \int_{t_0}^{t_1} dt_2 \int_{t_0}^{t_2} dt_3 \nonumber \\
&& \qquad\quad \times\, H(t_1) H(t_2) H(t_3) \; + \; \ldots ,
\end{eqnarray}
commonly referred to as the Dyson series. Here, the proposed problem can be stated as follows: given a known set of discrete data $\left\{ \rho_0, \rho(t_i) \right\}_{i=1}^{N_1}$ or $\left\{ U(t_j,t_0) \right\}_{j=1}^{N_2}$, find the continuous function $U(t, t_0)$.

By including the time-ordering operator $\mathcal{T}$, Eq.~\eqref{eq:dyson} can be written as
\begin{equation}
  \label{eq:unitary-evolution}
  U(t, t_0) = \mathcal{T} \exp \left( -\frac{i}{\hbar}\int_{t_0}^{t} H(\xi)\, \mathrm{d}\xi \right),
\end{equation}
where the time-ordering operator $\mathcal{T}$ ensures the proper ordering of Hamiltonian terms when the Hamiltonian at different times does not commute; that is, if $\comm{H(t)}{H(t')}\neq 0$ for any $t \neq t'$. For the sake of simplicity, we assume $t_0 = 0$, such that $U(t,t_0) = U(t)$.

It is important to note that, due to the presence of the time-ordering operator, Eq.~\eqref{eq:unitary-evolution} cannot, in general, be written as a simple matrix exponential such as $e^{-itH/\hbar}$, since the exact unitary arises from the infinite expansion. However, several approaches exist to evaluate Eq.~\eqref{eq:dyson}, including the Magnus expansion and the Lie--Trotter decomposition, among others~\cite{blanes2009, wiebe_simulating_2011}.  A brief description and numerical implementation of the Magnus and Trotter methods used in this work are provided in \ref{sec:app-integration}. These methods enter this work in two distinct roles. First, they provide the reference unitaries against which every model is evaluated, at all system sizes. Second, for the 7- and 8-qubit models they act as the propagation layer that maps the effective Hamiltonian predicted by the network to $U(t)$, as described in Sec.~\ref{sec:dl}. For systems of up to 6 qubits the network outputs the unitary directly, so no expansion is involved in the model itself. The two roles are discretized separately. Reference unitaries are computed with the Magnus expansion for the general Pauli and Ising systems, using 100 integration steps, and with the Trotterization for the $XYZ$ systems, using 25 steps. The propagation layer uses 50 steps during training and 10 at inference. The method used in each role is chosen independently, so a model propagating with Trotterization is also evaluated against Magnus-generated references.

\subsection{N-qubit systems}
\label{sec:n-qubit-systems}
Operators acting on an $N$-qubit Hilbert space can be spanned using the $N$-fold tensor product of the single-qubit Pauli basis $B = \{ I, \sigma_x, \sigma_y, \sigma_z \}$. Any Hamiltonian $H(t)$ acting on this space can therefore be expressed as a linear combination of the elements of $B^{\otimes N}$,
\begin{equation}
  \label{eq:hamiltonian-def}
  H(t) = \hbar \sum_{\alpha_1, \ldots, \alpha_N} c_{\alpha_1 \cdots \alpha_N}(t)\,
  \sigma_{\alpha_1} \otimes \sigma_{\alpha_2} \otimes \cdots \otimes \sigma_{\alpha_N},
\end{equation}
where each index $\alpha_j \in \{ I, x, y, z \}$ labels the Pauli operator acting on the $j$-th qubit, and the real-valued coefficients $c_{\alpha_1 \cdots \alpha_N}(t)$ carry the full time dependence. The sum runs over all $4^N$ combinations of the indices, i.e., over the complete set of $N$-qubit Pauli strings. For brevity, we collect the indices into a multi-index $\vec{\alpha} \equiv (\alpha_1, \ldots, \alpha_N)$, so that $c_{\vec{\alpha}}(t) \equiv c_{\alpha_1 \cdots \alpha_N}(t)$ and $\sigma_{\vec{\alpha}} \equiv \sigma_{\alpha_1} \otimes \cdots \otimes \sigma_{\alpha_N}$.

We consider three types of Hamiltonians across different many-body configurations. For systems with up to 6-qubits, we employ the general Hamiltonian of Eq.~(\ref{eq:hamiltonian-def}), which includes all possible Pauli components. For systems with 7- or 8-qubits, we restrict the Hamiltonian to two reduced nearest-neighbor models: an Ising model and an $XYZ$ Heisenberg model. These are explicitly given by
\begin{eqnarray}
H_{\mathrm{general}}(t)
&=&
\hbar \sum_{\alpha_1,\ldots,\alpha_N}
c_{\alpha_1 \cdots \alpha_N}(t)
\,
\sigma_{\alpha_1}\otimes\cdots\otimes\sigma_{\alpha_N},
\\
H_{\mathrm{Ising}}(t)
&=&
\hbar \left[ \sum_{j=1}^{N-1}
J_j(t)\,
\sigma_j^z \sigma_{j+1}^z
+
\sum_{j=1}^{N}
h_j(t)\,
\sigma_j^z \right],
\\
H_{XYZ}(t) &=& \hbar \sum_{j=1}^{N-1} \left[
J_j^x(t)\, \sigma_j^x \sigma_{j+1}^x
+ J_j^y(t)\, \sigma_j^y \sigma_{j+1}^y \right. \nonumber \\
&& \left. + J_j^z(t)\, \sigma_j^z \sigma_{j+1}^z \right].
\end{eqnarray}

Here $\sigma_j^{x,y,z}$ denote the Pauli operators acting on the $j$-th qubit. In the Ising model, $J_j(t)$ is the time-dependent coupling strength between neighboring qubits $j$ and $j+1$, while $h_j(t)$ is the time-dependent longitudinal field acting on qubit $j$. In the $XYZ$ Heisenberg model, $J_j^x(t)$, $J_j^y(t)$, and $J_j^z(t)$ are the time-dependent nearest-neighbor exchange couplings along the $x$, $y$, and $z$ directions, respectively. All coupling and field coefficients are real-valued functions of time.

The choice of a fully non-zero Pauli Hamiltonian for systems up to 6-qubits is motivated by its generality: since any $N$-qubit Hamiltonian can be expressed as a linear combination of all Pauli tensor products, a model capable of reproducing the dynamics generated by such a Hamiltonian should, in principle, be capable of handling any sparser Hamiltonian as a special case. This basis is used precisely because no structural prior on $H(t)$ is assumed, consistent with the setting described in Sec.~\ref{sec:intro}. When the interaction structure is known in advance, it can instead be imposed directly on the ansatz, which is what is done for the 7- and 8-qubit systems, where the effective Hamiltonian is restricted to nearest-neighbor form. The two choices therefore reflect how much prior structural information is assumed, rather than a limitation tied to system size. Both strategies are in fact applicable over a common range of system sizes, as shown in \ref{sec:app4}, where they are trained on the same Hamiltonian family over the overlapping range $N = 2$ to $6$. Imposing a known structure is also what makes the larger systems computationally tractable, since the number of coefficients to be learned then grows linearly rather than exponentially with $N$. Both reduced models are, in addition, directly relevant to physically realizable systems. The nearest-neighbor Ising model with longitudinal couplings and fields describes the effective spin dynamics in trapped-ion quantum simulators~\cite{Kim2011}, superconducting qubit architectures~\cite{Wu2016}, and quantum annealing devices such as D-Wave processors, where a time-dependent Hamiltonian drives the system toward the ground state of a programmable Ising problem~\cite{Kadowaki1998}. The nearest-neighbor $XYZ$ Heisenberg model describes anisotropic exchange interactions present in magnetic materials~\cite{Coldea2001}, has been experimentally realized in trapped-ion chains~\cite{Kim2012, Grass2014}, and serves as the effective low-energy Hamiltonian in silicon spin-qubit platforms~\cite{Zajac2018} and nuclear magnetic resonance quantum processors~\cite{Zhang2005}.

Furthermore, for all cases the time-dependent coefficients take the form
\begin{equation}
\label{eq:coeff-form}
c_{\vec{\alpha}}(t) = \frac{A_{\vec{\alpha}} \sin(\omega_{\vec{\alpha}}\, t + \phi_{\vec{\alpha}})}{\mathcal{N}_N},
\end{equation}
where $A_{\vec{\alpha}}$, $\omega_{\vec{\alpha}}$, and $\phi_{\vec{\alpha}}$ are the amplitude, angular frequency, and phase shift of each component, respectively, and $\mathcal{N}_N$ is a system-size-dependent normalization factor introduced to control the operator norm of the Hamiltonian ensuring the convergence of the Magnus expansion. The parameters are independently drawn from uniform distributions and, once fixed for a given realization, remain constant throughout training. For the general Pauli and Ising Hamiltonians we take $\omega_{\vec{\alpha}} \in [0, 8\pi)$, $\phi_{\vec{\alpha}} \in [0, 2\pi)$ and $A_{\vec{\alpha}} \in [0, 1)$. For the $XYZ$ Heisenberg systems the frequencies are drawn from $\omega \in [0, 2\pi)$, with the same phase range and amplitudes in $[0.5, 1.5)$, so that no bond is left close to decoupled. Each coefficient therefore completes up to four oscillations within the unit time interval in the first case and one in the second, and $U(t)$ inherits this oscillatory content. This sets the difficulty of the interpolation task and motivates the range of time steps $\Delta t$ considered.

The number of independent coefficients depends on the Hamiltonian structure: $4^N$ for the general Pauli case, $2N-1$ for the Ising model, and $3(N-1)$ for the $XYZ$ model. The normalization factors are
\begin{equation}
\mathcal{N}_2 = 2,\quad \mathcal{N}_3 = 3,\quad \mathcal{N}_4 = 4,\quad \mathcal{N}_5 = 16,\quad \mathcal{N}_6 = 24,
\end{equation}
for the general Pauli Hamiltonian, $\mathcal{N}_7 = \mathcal{N}_8 = 1$ for the Ising systems, where no normalization is needed since the nearest-neighbor Hamiltonian contains far fewer terms than the general case, and $\mathcal{N}_7 = \mathcal{N}_8 = 3$ for the $XYZ$ systems. With these choices, the sufficient convergence condition $\int_0^T \|H(t)\|_{\mathrm{s}}\,dt < \pi$ is satisfied in every configuration considered, with values of the integral between $1.0$ and $3.0$. In the general Pauli case the coefficient associated with the global identity operator is set to zero, ensuring that the Hamiltonian is traceless. The reduced models contain no identity component by construction.

The use of randomly sampled parameters allows the generation of a diverse ensemble of time-dependent Hamiltonians, preventing overfitting to specific dynamical patterns and enabling a systematic assessment of the model's ability to generalize across a broad class of physically valid quantum systems.

\section{Model description}
\label{sec:dl}
The rationale for employing PINNs lies in their capability to incorporate physical and mathematical constraints directly into the training process. PINNs have demonstrated outstanding performance in solving nonlinear differential equations and in modeling complex dynamical systems \cite{raissi2019, karniadakis2021}. Furthermore, neural networks consistently exhibit advantages in interpolation accuracy and adaptability to complex boundary and initial condition constraints.

\begin{figure*}[tp]
  \centering
  \includegraphics[width=0.99\textwidth]{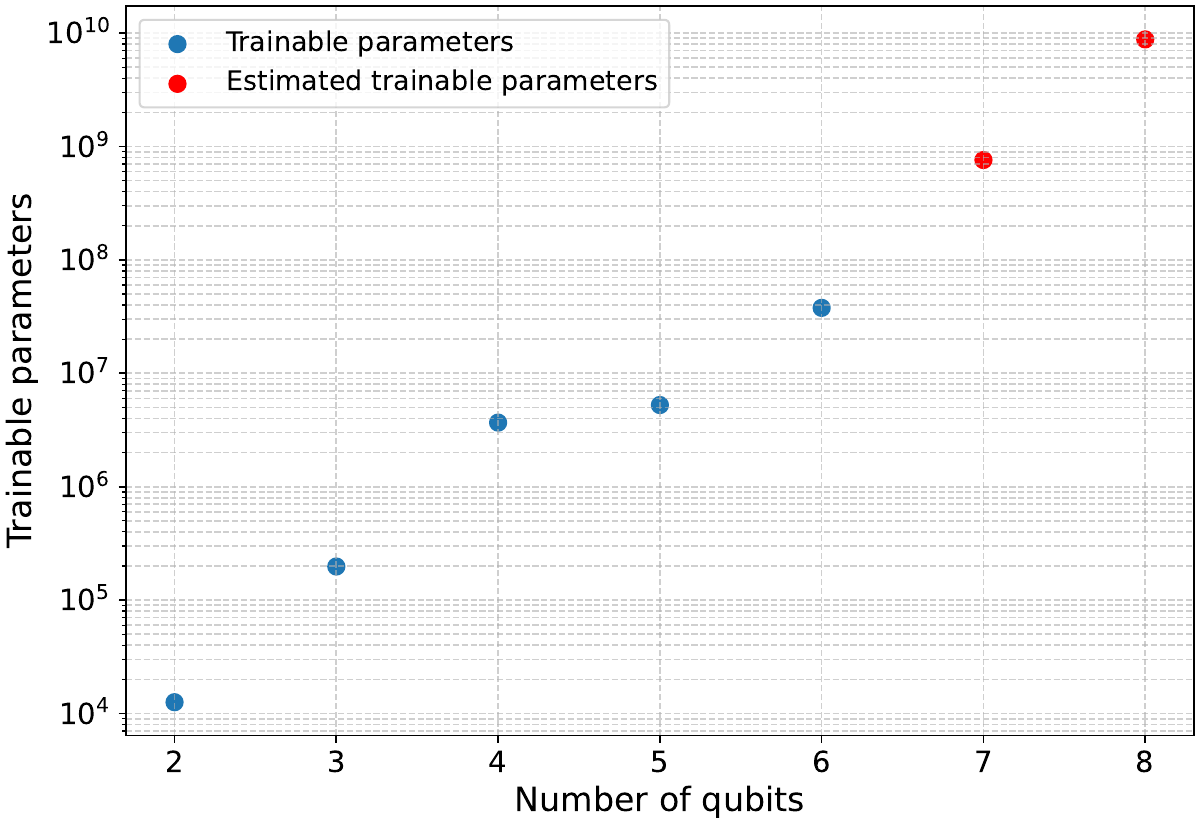}
  \caption{\label{fig:scalability}Amount of trainable parameters as a function of the number $N$ of qubits. In cases with up to 6 qubits, this relationship is outlined. For systems containing 7 and 8 qubits, we have provided the estimated number of trainable parameters required to apply the same strategy to determine the unitary evolution operator for those systems.}
\end{figure*}

We employ a model that generates a unitary complex-valued matrix from a scalar input. The dimensionality of these matrices increases exponentially with the number of qubits as $4^N$. Figure~\ref{fig:scalability} shows the number of trainable parameters for models up to 6-qubits, as well as the estimated parameter counts required for 7- and 8-qubit systems. Since the optimization process becomes computationally intractable at larger scales, an alternative strategy is adopted for these higher-dimensional cases. Rather than merely constraining the Hamiltonian to a reduced set of components, the model is constructed to estimate an effective time-dependent set of coefficients defining the effective Hamiltonian of the system $H_{\mathrm{eff}}(t)$. Subsequently, with a model capable of generating $H_{\mathrm{eff}}(t)$, we can use it to compute the time-evolution operator through (i) the Magnus expansion or (ii) the  Trotterization. This approach reduces the number of trainable parameters to approximately $10^3$.

\begin{figure*}[p]
  \centering

  \includegraphics[height=0.365\textheight]{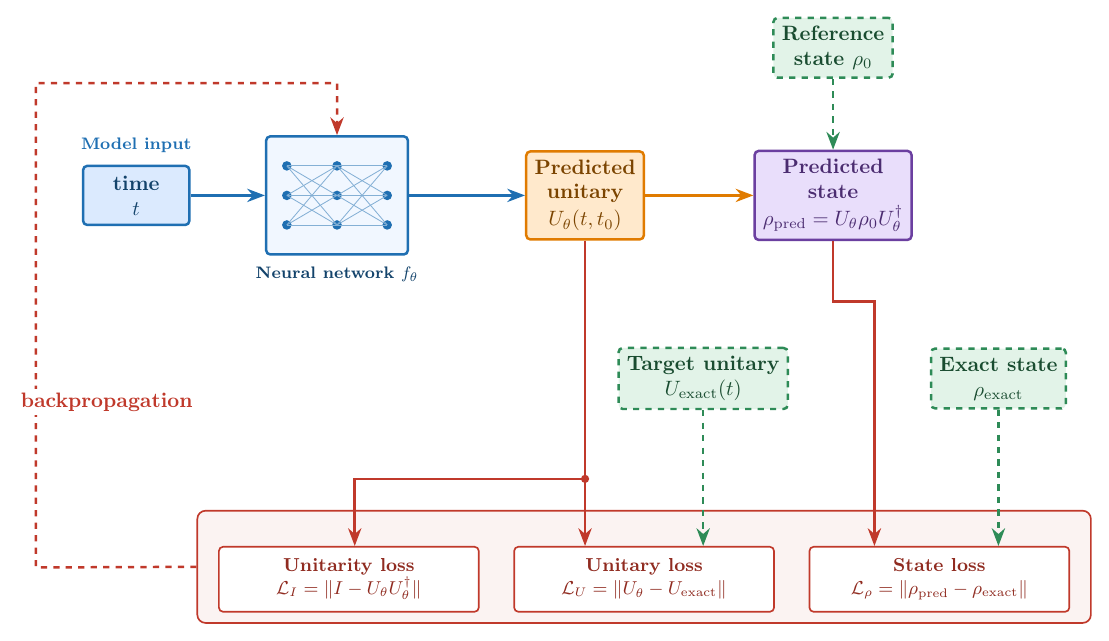}

  \vspace{1mm}
  {\small (a) PINN model for 2 to 6 qubits.}

  \vspace{5mm}

  \includegraphics[height=0.365\textheight]{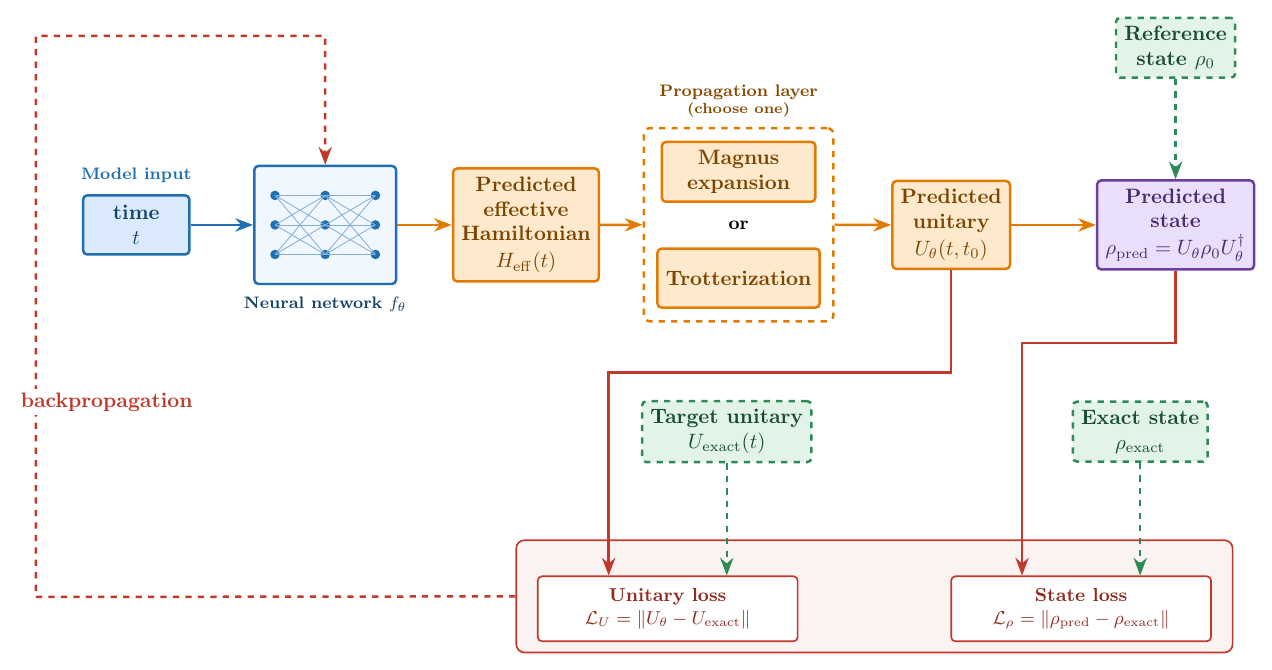}

  \vspace{1mm}
  {\small (b) PINN model for 7 and 8 qubits.}

  \caption{\textbf{The two PINN architectures used to emulate the time evolution of the isolated multi-qubit systems.} In both cases $f_\theta$ takes the time $t$ as its only input and is trained by minimizing the loss. \textbf{(a)} For 2 to 6 qubits, $f_\theta$ outputs the unitary $U_\theta(t,t_0)$ directly, and the loss combines $\mathcal{L}_I$, $\mathcal{L}_U$ and $\mathcal{L}_\rho$. \textbf{(b)} For 7 and 8 qubits, $f_\theta$ predicts an effective Hamiltonian $H_{\mathrm{eff}}(t)$ that a propagation layer (Magnus expansion or Trotterization) maps to $U_\theta(t,t_0)$, unitary by construction, so $\mathcal{L}_I$ is dropped. The $\mathcal{L}_U$ term, which requires the target unitaries to be known, enters both architectures and can be removed without significant loss of accuracy, as shown in \ref{sec:app4}.} \emph{Color code:} blue, model input and outputs; orange, quantities learned by the network; purple, predicted state; dashed green, quantities not produced by the network; red, loss terms and the optimization path.
  \label{fig:models-description}
\end{figure*}

\subsection{Architecture description}
For the model architecture, we employ a straightforward approach consisting of a sequence of fully connected layers designed to match the number of parameters required for the output matrix. The hyperbolic tangent function is used as the activation function. This architecture can generate arbitrarily complex-valued matrices, ensuring that the model can accurately represent any matrix form of the time-evolution operator. Figure~\ref{fig:models-description} schematically illustrates the two modeling strategies adopted for the time-evolution operator. A detailed description of the architectures used in the different models is provided in \ref{sec:app1}.

\subsection{Loss function and data generation}
\label{sec:loss-function}
To train the model, a valid dataset is required. The dataset generation process is inspired by the idea that it can be interpreted as the outcome of multiple QPT experiments performed at different times~\cite{huang2025}. If the goal is to perform a quantum control task, the dataset will consist of a set of quantum states forming a trajectory and target unitary operators applied at specific times $t$. However, for practical purposes, the dataset is generated numerically by computing the system dynamics using Eqs.~\eqref{eq:state-evolution} and \eqref{eq:unitary-evolution}.

Employing the Magnus expansion or the Trotterization method increases the computational cost, as these calculations must be performed at each training iteration to update the model parameters. To mitigate this, the Magnus expansion is truncated at second order. For both, the Magnus and Trotterization approaches, we use 50 integration steps per model query, i.e., each time the model produces an output.

Characterizing $U(t)$ at a single time by quantum process tomography is expensive. In the standard tomographic setting, reconstructing a general process on $N$ qubits requires $\mathcal{O}(16^N)$ experimental configurations, and even when unitarity is assumed a priori the operator still carries $4^N - 1$ real parameters. Reducing this cost is an active line of research~\cite{huang2025}, complementary to the direction taken here. Since each characterization constrains $U(t)$ at one instant only, resolving a continuous time dependence in this way would require an impractical number of such experiments. The approach proposed here does not reduce the cost of an individual characterization, but the number of characterizations needed, by interpolating the evolution between the few times at which data are available. The study focuses on the time interval $t \in [0, 1]$, employing discrete time grids with step sizes of $\Delta t = 0.1, 0.25, 0.5$, and $1.0$. This procedure yields a comprehensive dataset for the selected time values, essential for effective model training. The supervision data consist of two sets of different size. The first is a set of state pairs $\left\{ \left( \rho_0^i, \tau_i, \rho_{\tau_i}^i \right) \right\}_{i=1}^{N_{\rho}}$, with $N_{\rho} = 11000$, whose evolved states satisfy $\rho_{\tau_i}^i = U(\tau_i,0)\, \rho_0^i\, U^{\dagger}(\tau_i,0)$. The second is the set of evolution operators $\left\{ \left( t_j, U_{t_j} \right) \right\}_{j=1}^{N_U}$ known at the sampled times. The two differ in what they cost to obtain: a state pair requires a single preparation and measurement, whereas each $U_{t_j}$ requires a full characterization at that time. Accordingly, $N_U$ is the number of points of the time grid and is much smaller than $N_{\rho}$. The evolution times $\tau_i \in [0,1]$ are sampled from uniform time grids corresponding to different choices of the time step $\Delta t$. The initial states $\rho_0^i$ are random mixed states sampled from the Hilbert--Schmidt measure~\cite{zyczkowski2001}: a complex matrix $G$ with independent Gaussian real and imaginary entries is drawn, and the state is obtained as $\rho_0 = G G^\dagger / \mathrm{Tr}(G G^\dagger)$, which is Hermitian, positive semi-definite, and normalized by construction. This dataset size was not optimized; \ref{sec:app4} reports comparable accuracy with one to two orders of magnitude fewer samples.

The loss function is defined as an objective function whose minimization yields the optimal parameters of the deep learning model, such that the resulting model accurately describes the dynamics of the physical system under consideration. It is constructed to explicitly incorporate the known physical constraints governing the system dynamics, as well as the available supervision data. In particular, it enforces consistency with unitary quantum evolution, agreement with known unitary operators when available, and compliance with the von Neumann equation~\eqref{eq:von-neumann-eq}. The total loss function is defined as

\begin{eqnarray}
\label{eq:loss}
\mathcal{L} &=& \mathcal{L}_{\rho} + \mathcal{L}_{U} + \mathcal{L}_{I}, \qquad \mathrm{with} \nonumber \\[0.4em]
\mathcal{L}_{\rho} &=& \frac{1}{N_{\rho}} \sum_{i=1}^{N_{\rho}} \big\Vert \rho_{\tau_i}^i - U_{\theta}(\tau_i)\, \rho_0^i\, U_{\theta}^{\dagger}(\tau_i) \big\Vert , \nonumber \\
\mathcal{L}_{U} &=& \frac{1}{N_U} \sum_{j=1}^{N_U} \big\Vert U_{\theta}(t_j) - U_{t_j} \big\Vert , \nonumber \\
\mathcal{L}_{I} &=& \frac{1}{N_f} \sum_{k=1}^{N_f} \big\Vert \mathbb{1} - U_{\theta}(t_k)\, U_{\theta}^{\dagger}(t_k) \big\Vert .
\end{eqnarray}

where $\Vert A \Vert = \sum_{ij} |a_{ij}|$ denotes the element-wise $\ell_1$ matrix norm for any matrix $A = [a_{ij}]$, and $U_{\theta}(t)$ represents a neural network parametrization of the time-evolution operator with trainable weights $\theta = \{ \theta_i \}$. Physics-Informed Neural Networks employ such neural representations to approximate unknown physical quantities while enforcing governing equations and constraints directly through the loss function \cite{raissi2019}.

Each term is averaged over its own data, since the three are supported on sets of different size: $N_{\rho}$ state pairs, $N_U$ known evolution operators, and $N_f$ temporal collocation points used to enforce unitarity across the time domain. The collocation points carry no measured data and can therefore be chosen freely. The first term of the loss, $\mathcal{L}_{\rho}$, enforces accurate quantum state evolution under the predicted unitary operator $U_{\theta}$. The second term, $\mathcal{L}_{U}$, promotes consistency between the predicted unitaries and known target unitaries, when such information is available. The third term, $\mathcal{L}_{I}$, explicitly imposes the physical constraint of unitarity by penalizing deviations from $U_{\theta}(t) U_{\theta}^{\dagger}(t) = \mathbb{1}$ at arbitrary times.

For the 7- and 8-qubit systems, only the first two terms are employed. This choice is motivated by the fact that both the Magnus expansion and the Trotterization intrinsically preserve unitarity, rendering an explicit unitarity penalty unnecessary in these cases. During training, both methods use $r = 50$ integration steps. All results presented in the main text are obtained with the full loss function of Eq.~\eqref{eq:loss}, the most favorable training scenario, in which both the evolved states and the target unitary operators are available. The unitary term is the more demanding one experimentally, since it presupposes knowledge of the evolution operator at the sampled times. In \ref{sec:app4} we show that it can be removed without significant loss of accuracy whenever the temporal sampling is sufficient, so that the model can be trained from state data alone.

\subsection{Test and training}
\label{sec:test-train}
The AdamW optimizer is used for training, initialized with a learning rate of $10^{-3}$. To enhance convergence stability, a learning-rate scheduler decreases the rate by a factor of $10^{-1}$ every 300 epochs. The training process spans $10^3$ epochs, each consisting of 100 data batches with a batch size of 10. Each data set contains $11000$ uniformly distributed tuples in the specified time intervals. In addition, the data batches used to train the models for 7- and 8-qubit systems are reduced to 10 data batches per epoch. All models presented in this work were trained using an NVIDIA Quadro P6000 GPU. For all model sizes, we observe that datasets with a larger $\Delta t$ yield a more pronounced decrease in the training loss, since the data are less constrained and contain fewer distinct evolution times. We emphasize, however, that a lower training loss does not by itself imply better interpolation: our goal is to predict the unitary at times not seen during training, and a more constrained dataset (smaller $\Delta t$) consistently yields better interpolation accuracy, as quantified by the test-set metrics reported in Sec.~\ref{sec:results}. The full training-loss curves for all system sizes and training configurations are available in the public repository accompanying this work.

\section{Results}
\label{sec:results}
To assess the model’s performance over the entire time domain we compute 100 target unitary matrices. These matrices are uniformly spaced in time and include values not present in the training set, thereby allowing us to rigorously evaluate the model's interpolation capability. To quantify accuracy, we adopt the gate fidelity~\cite{cabrera2011}, defined as
\begin{equation}
  \label{eq:metric-fidelity}
  F(U(t), U_{\theta_{\mathrm{opt}}}(t)) = \left| \frac{1}{d} \Tr \left[ U^{\dagger}(t) U_{\theta_{\mathrm{opt}}}(t) \right] \right|^2,
\end{equation}
where $\theta_{\mathrm{opt}}$ are the optimal weights of the Neural Network found during training, and $d = 2^N$ denotes the Hilbert-space dimension for an $N$-qubit system. The fidelity ranges from 0 to 1, with $F = 1$ only when the predicted operator $U_{\theta_{\mathrm{opt}}}(t)$ coincides with the target $U(t)$ at time $t$. This is a natural choice for comparing unitary matrices, as shown in \ref{sec:app3}.

For evaluation, each trained model generates unitary matrices at the same 100 time points. These predicted matrices are then compared one-by-one with the target unitaries, yielding a fidelity function $F(t)$. Since multiple models for each system size are trained with different values of $\Delta t$, four fidelity curves $F(t)$ are obtained for each system size.

The loss function additionally enforces unitarity across the time domain, ensuring that the model’s predictions remain close to unitary transformations throughout training. To further minimize numerical deviations, we can compute the closest unitary matrix to each predicted output via Singular Value Decomposition (SVD), redefining the estimated unitary as
\begin{equation}
  \label{eq:error-correction}
  U_{\theta_{\mathrm{opt}}}(t) = U V_{\mathrm{h}},
\end{equation}
where $U$ and $V_{\mathrm{h}}$ arise from the SVD of the model’s predicted matrix. The following results compare the performance of the different models with and without post-processing, showing that high fidelity is preserved even in the absence of the closest-unitary correction.

\begin{figure*}[tp]
  \centering
  \includegraphics[width=0.99\textwidth]{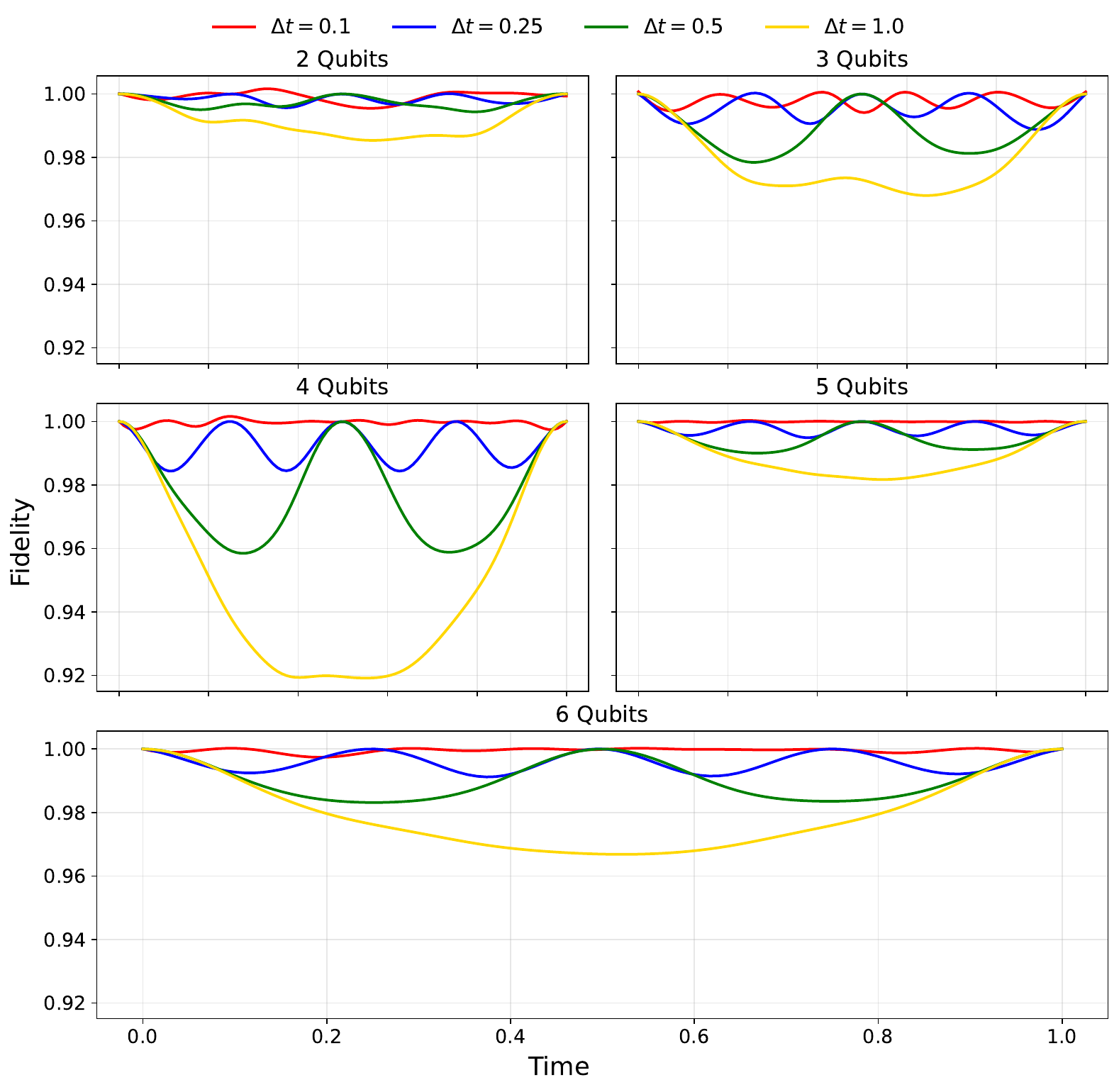}
  \caption{\label{fig:fid_results_2to6_without}Fidelity of the predicted unitary evolution for model2–model6, before SVD-based post-processing error correction.}
\end{figure*}

\begin{figure*}[tp]
  \centering
  \includegraphics[width=0.99\textwidth]{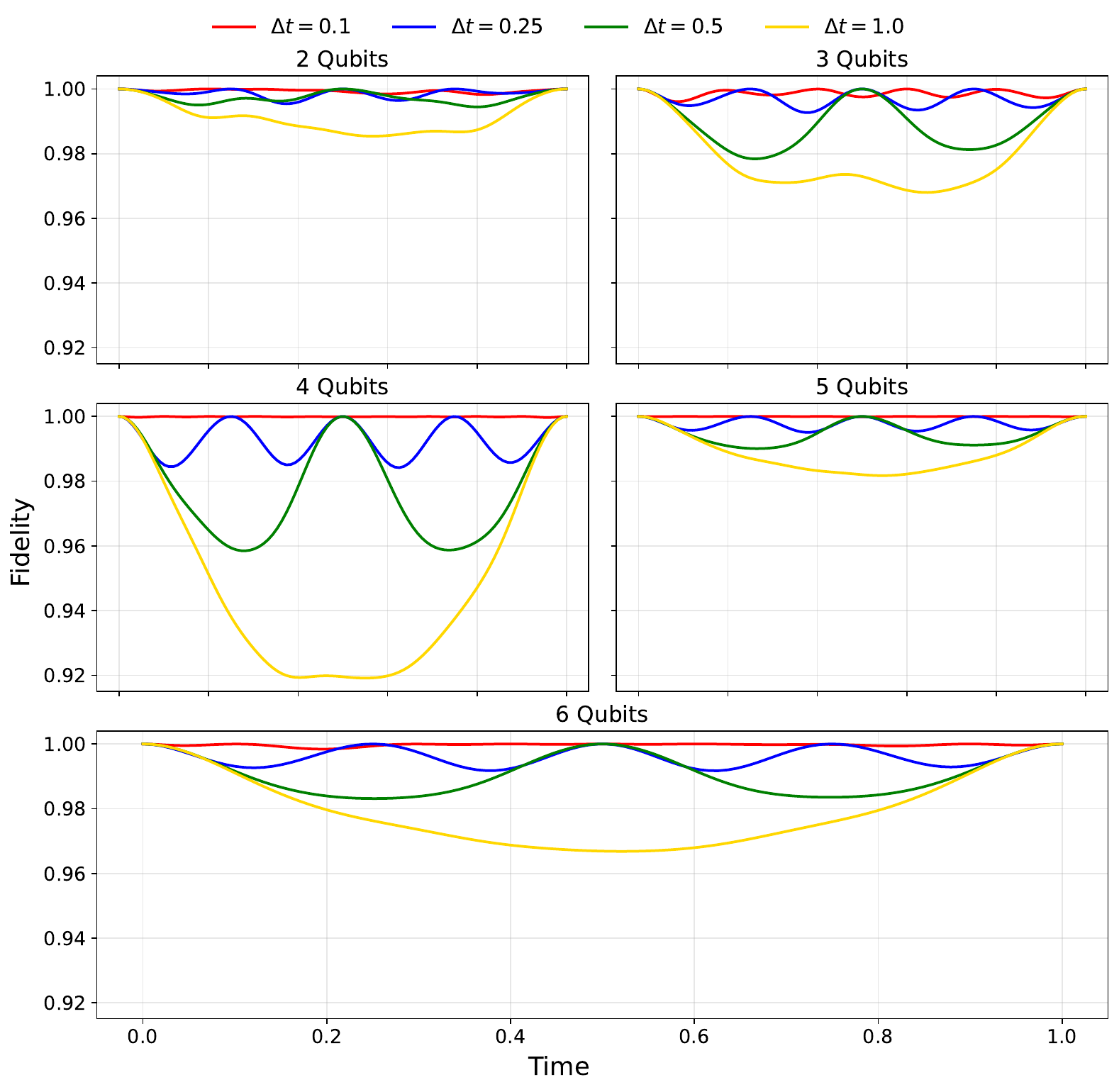}
  \caption{\label{fig:fid_results_2to6_with}Fidelity of the predicted unitary evolution for model2–model6, after SVD-based post-processing error correction.}
\end{figure*}

The results in Figs.~\ref{fig:fid_results_2to6_without} and~\ref{fig:fid_results_2to6_with} summarize the fidelity performance for systems up to 6-qubits, respectively before and after the error correction via \eqref{eq:error-correction}. As expected, datasets with smaller time steps ($\Delta t$) achieve the highest fidelities due to their finer temporal resolution. Nevertheless, even for larger $\Delta t$, the reduction in fidelity remains minimal. This reflects the PINN's strong interpolation capability, which maintains an accurate approximation of the quantum dynamics even with coarse temporal discretization.

\begin{figure*}[tp]
  \centering
  \includegraphics[width=0.99\textwidth]{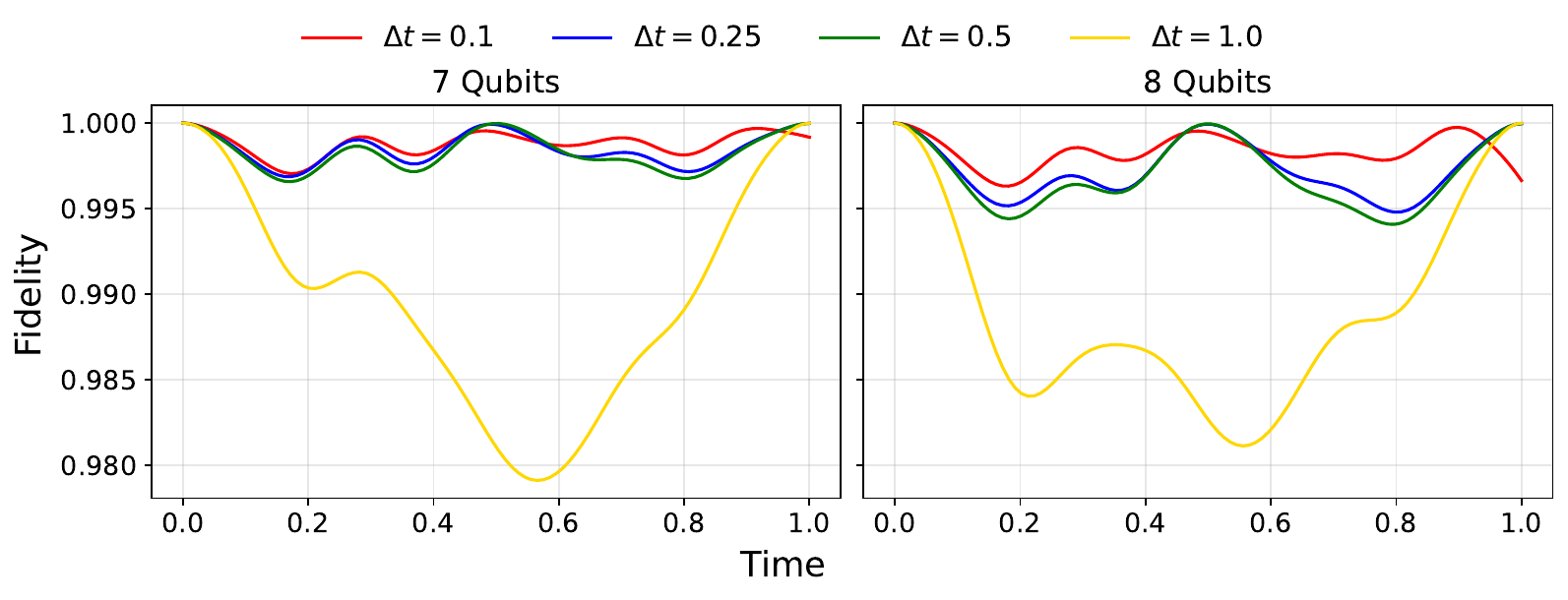}
  {\small (a)}

  \medskip

  \includegraphics[width=0.99\textwidth]{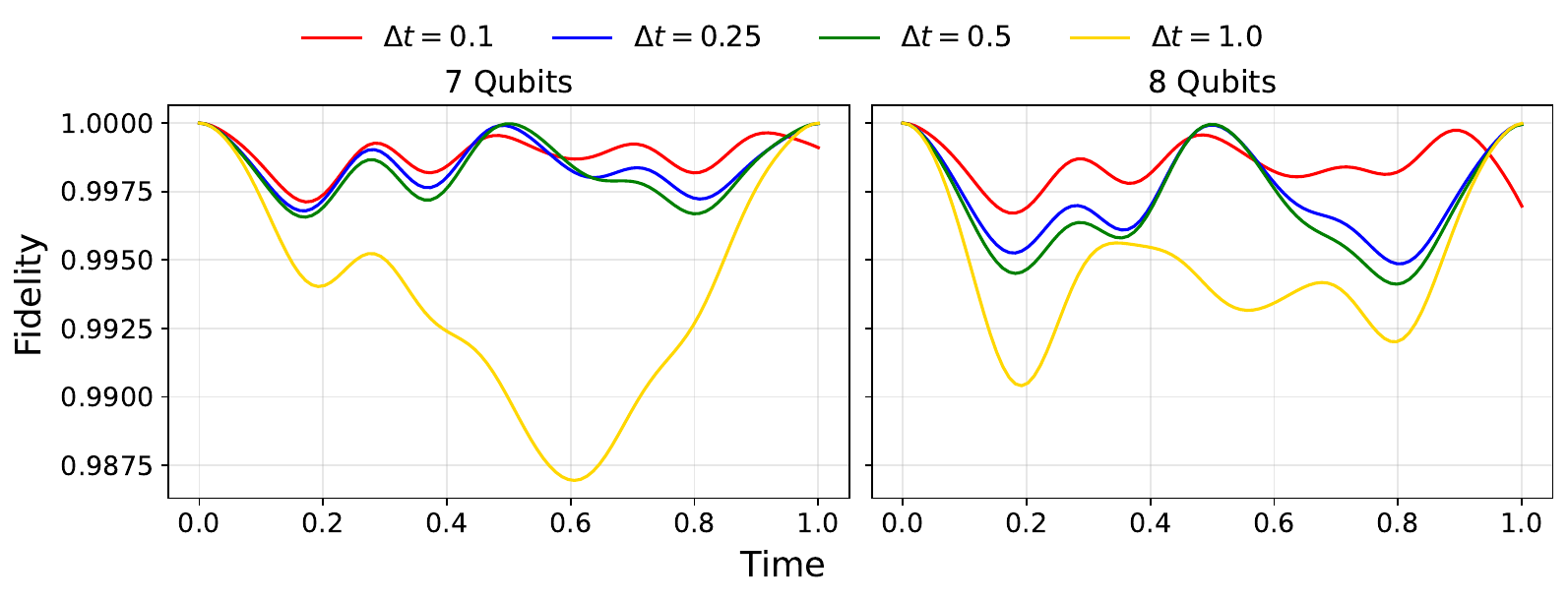}
  {\small (b)}

  \caption{\label{fig:fid78}Fidelity performance for 7- and 8-qubit systems under the four training setups considered for the Ising Hamiltonian system. (a) Trotter-based model. (b) Magnus-based model.}
\end{figure*}

\begin{figure*}[tp]
  \centering
  \includegraphics[width=0.99\textwidth]{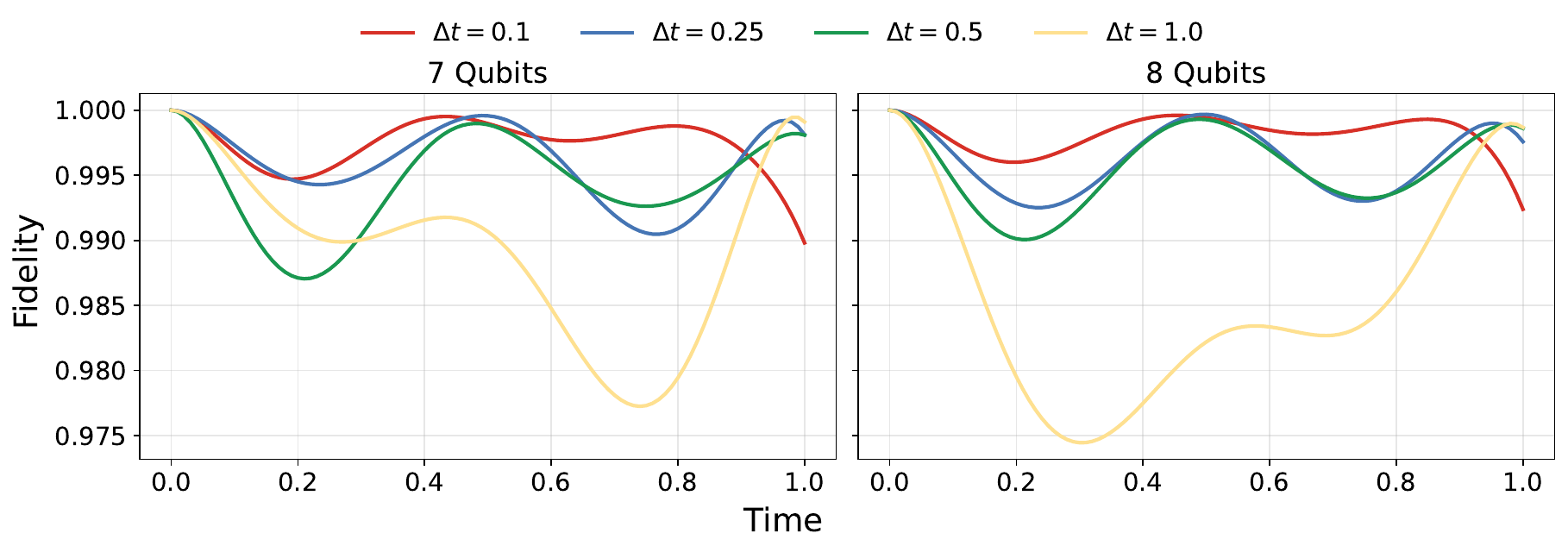}
  {\small (a)}

  \medskip

  \includegraphics[width=0.99\textwidth]{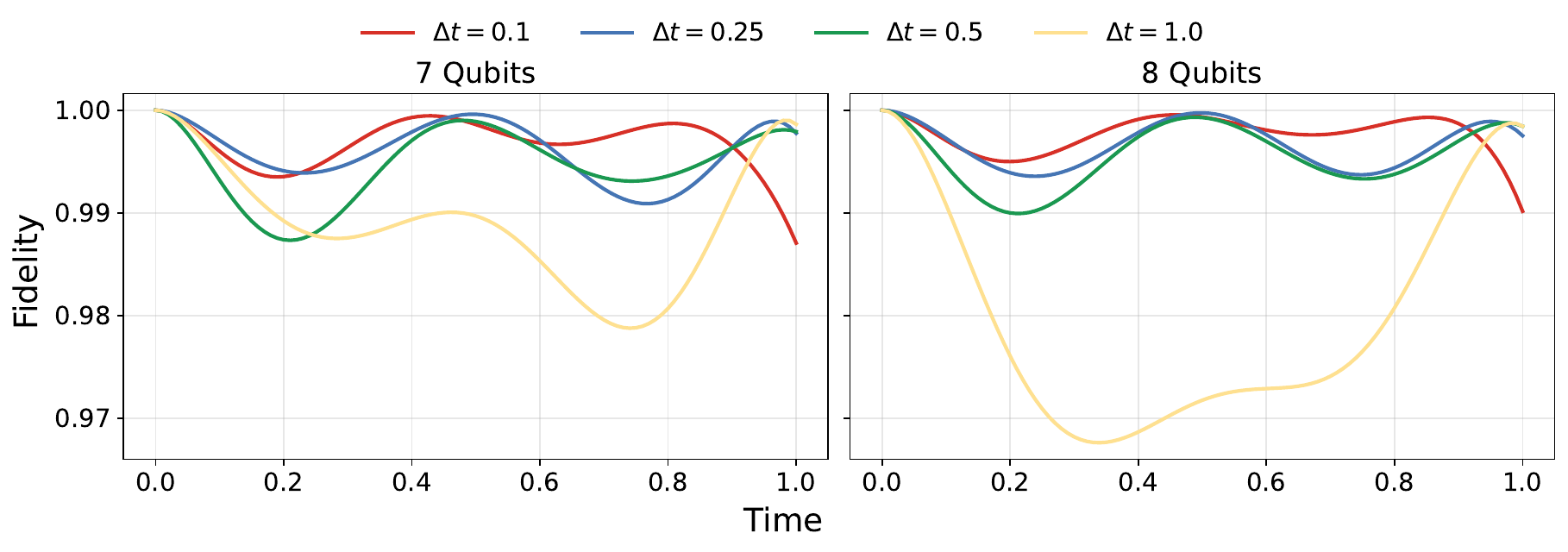}
  {\small (b)}

  \caption{\label{fig:fid78xyz}Fidelity performance for 7- and 8-qubit systems under the four training setups considered for the $XYZ$ Hamiltonian system. (a) Trotter-based model. (b) Magnus-based model.}
\end{figure*}

For the 7- and 8-qubit cases, the model follows a different strategy by estimating an effective Hamiltonian as an intermediate step to compute the corresponding unitary matrix. Since evolution can be achieved through Trotterization or Magnus expansion, both approaches were implemented and evaluated independently without SVD post-processing, as both methods guarantee unitarity. Figures~\ref{fig:fid78} and~\ref{fig:fid78xyz} show the fidelity results for these higher-dimensional systems across different training datasets, for the Ising and $XYZ$ Hamiltonians respectively. Consistent with the behavior observed in Figs.~\ref{fig:fid_results_2to6_without} and~\ref{fig:fid_results_2to6_with}, even for large values of $\Delta t$, the degradation in fidelity is minimal, demonstrating that the interpolation remains highly effective in all tested scenarios.

\begin{figure*}[tp]
\centering
\includegraphics[width=0.99\textwidth]{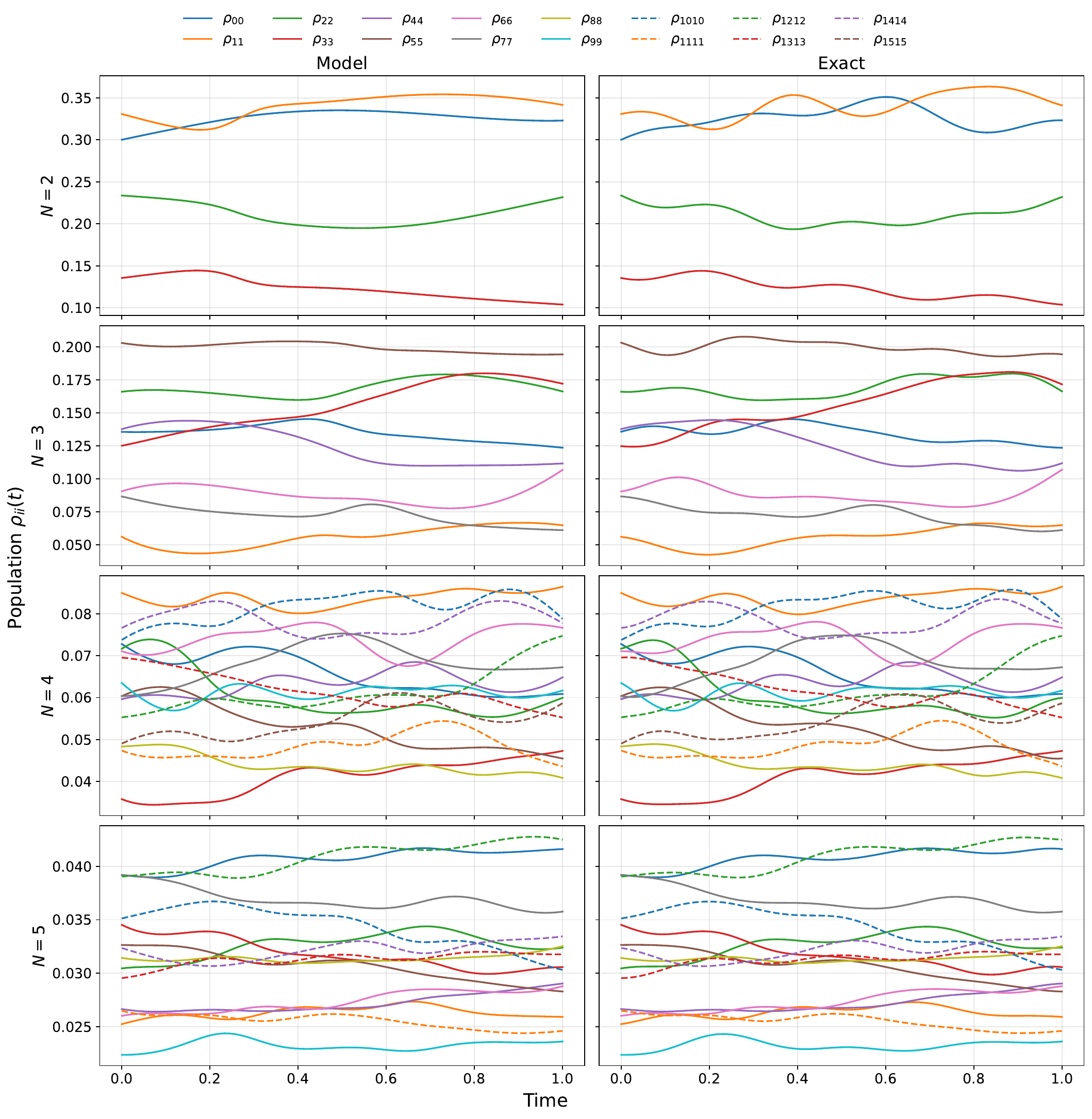}
\caption{
Diagonal elements $\rho_{ii}(t)$ of the evolved density matrix for systems of 2- to 5-qubits, comparing the model prediction (left column) against the exact unitary evolution (right column). 
Each curve corresponds to a different basis state population, for a single randomly sampled initial state $\rho_0$ and training time step $\Delta t = 0.1$. The physical system considered 
is described by the general Pauli Hamiltonian $H_{\mathrm{general}}(t)$, which includes all possible Pauli components in the $N$-qubit operator basis.
}
\label{fig:state-components-dynamics-2to5}
\end{figure*}

\begin{figure*}[tp]
\centering
\includegraphics[width=0.99\textwidth]{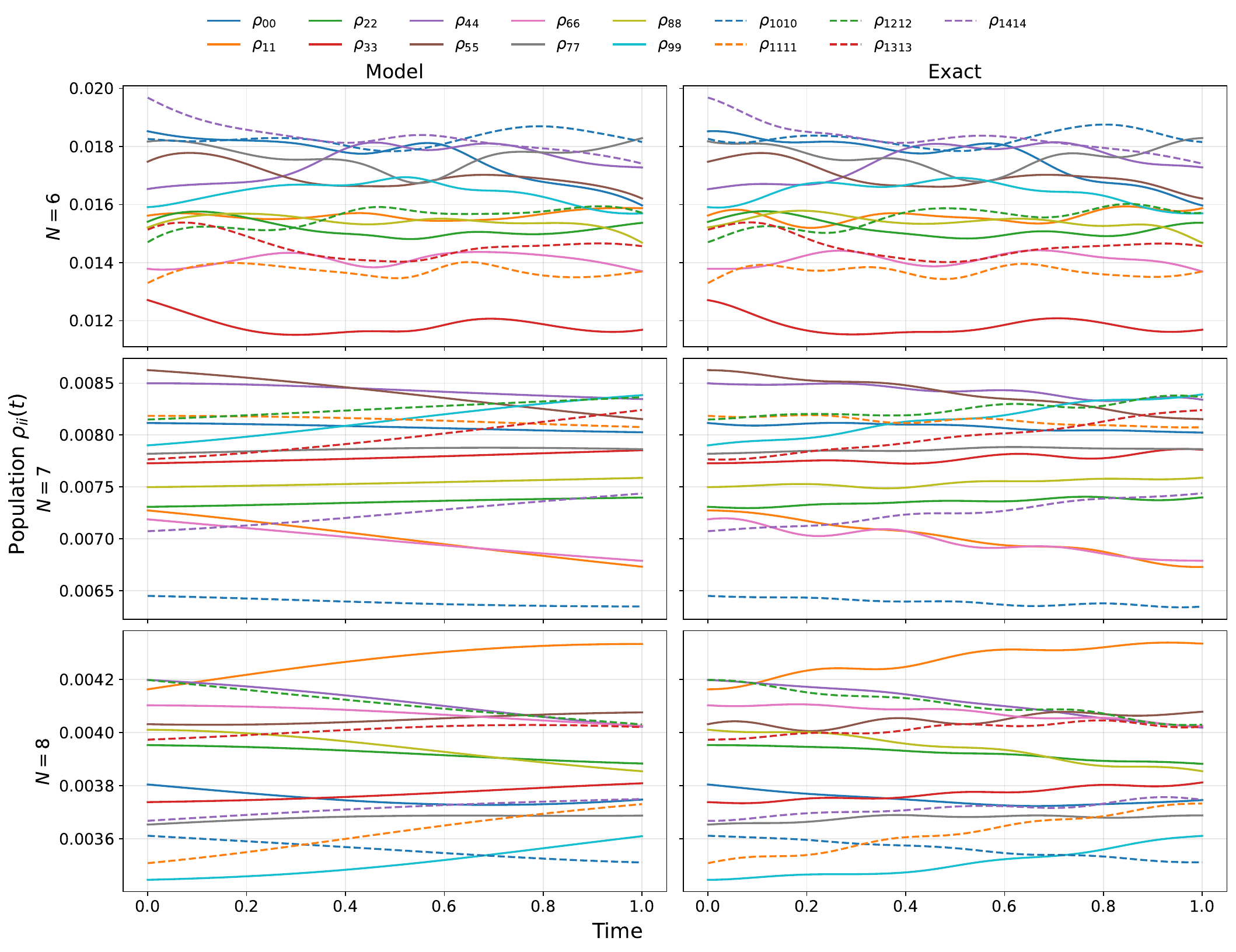}
\caption{
Same as Fig.~\ref{fig:state-components-dynamics-2to5} for systems of 6- to 8-qubits. For the 6-qubit case, the physical system is described by the general Pauli Hamiltonian
$H_{\mathrm{general}}(t)$. For the 7- and 8-qubit cases, the physical system is described by the nearest-neighbor Ising Hamiltonian $H_{\mathrm{Ising}}(t)$, for which the model predicts the coefficients of an effective Hamiltonian $H_\mathrm{eff}(t)$, from which the unitary is computed via the Magnus expansion.
}
\label{fig:state-components-dynamics-6to8}
\end{figure*}

To further validate the model's ability to reproduce quantum state dynamics, we evolve a randomly sampled initial state $\rho_0$ under two distinct time-evolution operators: (i) the exact unitary $U(t)$ computed via the second-order Magnus expansion, and (ii) the unitary $U_\theta(t)$ predicted by the trained model without SVD correction. At each time $t \in [0,1]$, we compute the evolved density matrix according to Eq.~\eqref{eq:state-evolution} and track the diagonal elements $\rho_{ii}(t)$, which correspond to the occupation probabilities of the computational basis states. Figures~\ref{fig:state-components-dynamics-2to5} and~\ref{fig:state-components-dynamics-6to8} show these populations for systems ranging from 2- to 8-qubits, comparing the model predictions against the exact reference. In all cases, the trajectories are in close agreement, with only minor deviations consistent with the gate fidelity values reported in the preceding sections.

\subsection{Statistical analysis}
To ensure that our evaluation did not rely on a single Hamiltonian realization per system size, we performed an additional statistical analysis to assess the model's robustness in interpolating unitary dynamics across diverse configurations. Three families of Hamiltonians were considered: the general Pauli Hamiltonian with all non-zero components (systems up to 6-qubits), the nearest-neighbor Ising model (7- and 8-qubits), and the nearest-neighbor $XYZ$ Heisenberg model (7- and 8-qubits). For the general Pauli case, we carried out 100 independent training runs per system size. Due to computational resource constraints, this analysis was limited to 50 independent runs for the 7- and 8-qubit Ising Hamiltonian, 50 independent runs for the 7-qubit $XYZ$ Hamiltonian, and 20 independent runs for the 8-qubit $XYZ$ Hamiltonian. In each run, the amplitudes, frequencies, and phases defining the Hamiltonian were independently sampled from uniform distributions according to Section~\ref{sec:n-qubit-systems}. This corresponds to $4^N - 1$ independent parameters for systems of up to 6 qubits, $2N - 1$ parameters for the Ising Hamiltonians, and $3(N-1)$ parameters for the $XYZ$ Hamiltonians. These parameters were used to construct the target unitary evolutions, generate the corresponding training datasets, and train the model under identical conditions. Applying the same fidelity-evaluation pipeline yielded a collection of fidelity curves $F^{(k)}(t)$, with $k = 1, \ldots, N_r$, where $N_r$ is the number of independent runs for each system size and Hamiltonian family. Each curve is obtained by evaluating the gate fidelity of Eq.~\eqref{eq:metric-fidelity} on a uniform grid of 100 points in $t \in [0,1]$, comparing the model prediction against the numerically computed target unitary. The mean fidelity $F_{\mu}(t)$ and the fidelity standard deviation $F_{\sigma}(t)$ were computed as
\begin{eqnarray}
  F_{\mu}(t) &=& \frac{1}{N_r} \sum_{k=1}^{N_r} F^{(k)}(t), \\
  F_{\sigma}(t) &=& \sqrt{ \frac{1}{N_r-1} \sum_{k=1}^{N_r} \left( F^{(k)}(t) - F_{\mu}(t) \right)^2 },
\end{eqnarray}
where $F^{(k)}(t)$ is the gate fidelity of the $k$-th run at time $t$.

\begin{figure*}[tp]
  \centering
  \includegraphics[width=0.99\textwidth]{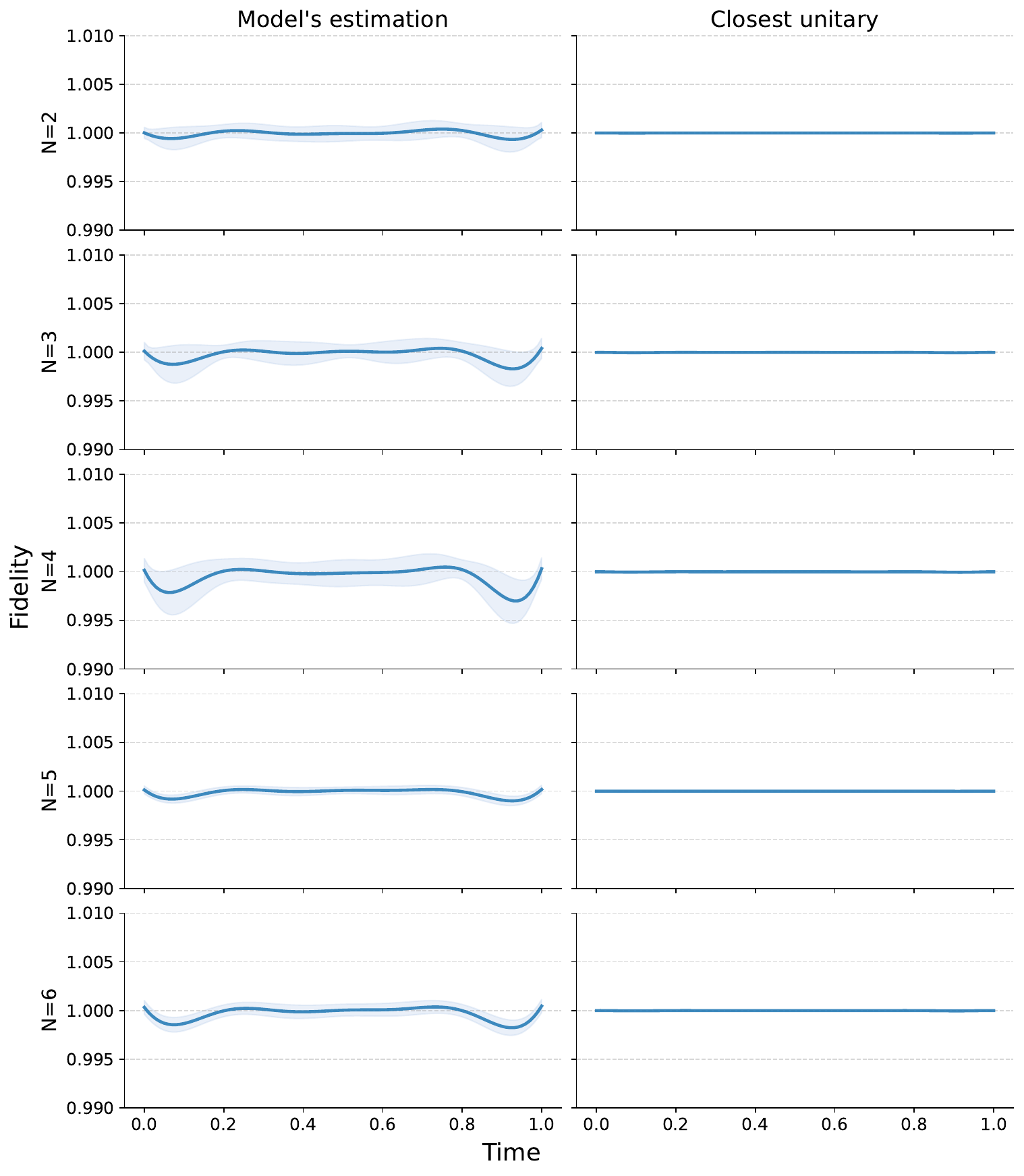}
  \caption{\label{fig:statistic-plot-to6}Mean fidelity and standard deviation across 100 independent training runs for systems up to 6 qubits. Left: raw model output vs.\ target unitary. Right: SVD-corrected output vs.\ target unitary.}
\end{figure*}

\begin{figure*}[tp]
  \centering
  \includegraphics[width=0.99\textwidth]{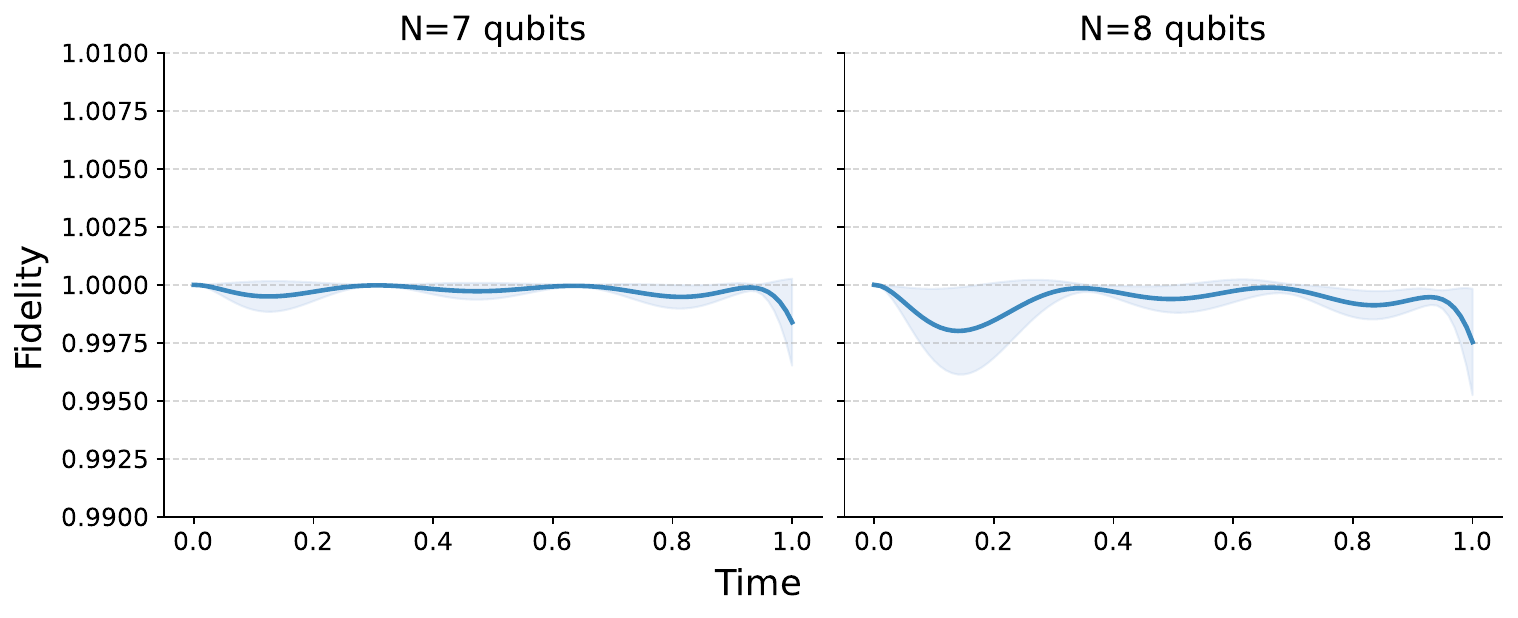}
  \caption{\label{fig:statistic-plot-78-ising}Same as Fig.~\ref{fig:statistic-plot-to6} for 7- and 8-qubit systems under the nearest-neighbor Ising Hamiltonian (50 independent runs).}
\end{figure*}

\begin{figure*}[tp]
  \centering
  \includegraphics[width=0.99\textwidth]{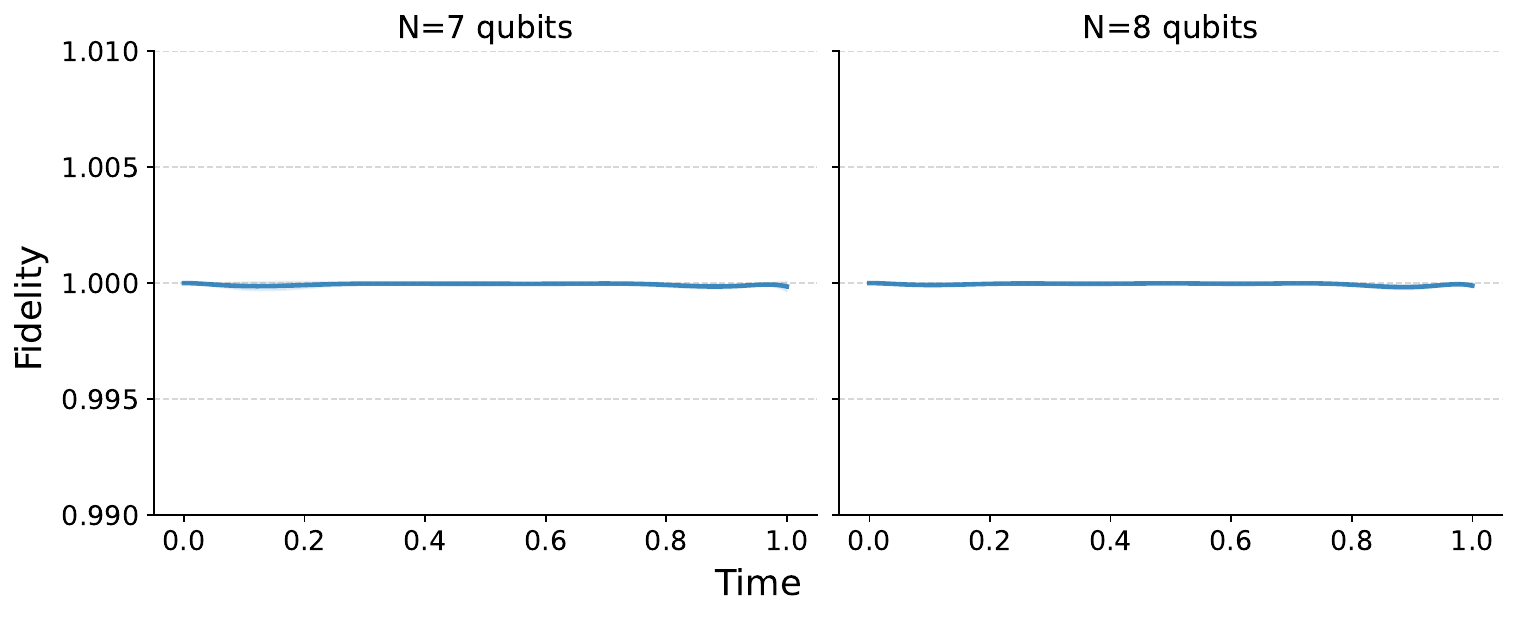}
  \caption{\label{fig:statistic-plot-78-xyz}Same as Fig.~\ref{fig:statistic-plot-to6} for 7- and 8-qubit systems under the nearest-neighbor $XYZ$ Heisenberg Hamiltonian (50 runs for 7 qubits, 20 runs for 8 qubits).}
\end{figure*}

Figures~\ref{fig:statistic-plot-to6}, \ref{fig:statistic-plot-78-ising}, and \ref{fig:statistic-plot-78-xyz} summarize the mean fidelity and standard deviation across all independent runs, providing a quantitative assessment of the variability in the model's performance. For each system size up to 6-qubits, two panels are shown: the first compares the raw model outputs with the target unitary evolutions, while the second reports the nearest-unitary matrices obtained through SVD-based post-processing. For 7- and 8-qubit systems, because unitarity is guaranteed by the Trotterization or Magnus expansion, SVD corrections are not required. Since these results closely match those obtained in the single-training analysis, we expect a similar statistical behavior for other values of $\Delta t$ when applying the same procedure.

Across all analyses, the standard deviation remains consistently low. Even without enforcing unitarity via SVD, the variability is small, and after error correction the associated uncertainty becomes negligible where it is not visually distinguishable in the figure. Moreover, no specific regions were identified where the model failed to interpolate successfully.

To extend our statistical analyses, we employ the trace distance as an additional figure of merit. For two density matrices $\rho$ and $\sigma$, the trace distance is defined as
\begin{equation}
\label{eq:trace-distance}
T(\rho, \sigma) = \frac{1}{2} \left\Vert \rho - \sigma \right\Vert_1 = \frac{1}{2} \sum_i |\lambda_i|,
\end{equation}
where $\lambda_i$ are the eigenvalues of the Hermitian operator $\rho - \sigma$, and $T \in [0,1]$. Following the same procedure as in the statistical analysis above, we evaluate the trace distance between the states evolved under the predicted and exact unitaries over an ensemble of $N_s = 50$ random initial states, on a uniform grid of $N_t = 100$ points in $t \in [0,1]$. The initial states are sampled from the Hilbert-Schmidt measure, as described in Sec.~\ref{sec:loss-function}.

For each qubit size, we compute the trace distances
\begin{equation}
  T_i(t) = T\left(\rho_i(t),\sigma_i(t)\right),
\end{equation}
where $\rho_i(t) = U_{\theta_{\mathrm{opt}}}(t)\, \rho_0^i\, U_{\theta_{\mathrm{opt}}}^{\dagger}(t)$ and $\sigma_i(t) = U(t)\, \rho_0^i\, U^{\dagger}(t)$ are the states evolved under the predicted and exact unitaries respectively, and $i=1,\ldots,N_s$ labels the random initial states. The reported value corresponds to the mean trace distance,
\begin{equation}
  T_{\mu}(t)= \frac{1}{N_s} \sum_{i=1}^{N_s} T_{i}(t),
\end{equation}
while the error bars indicate one standard deviation,
\begin{equation}
  T_{\sigma}(t) = \sqrt{ \frac{1}{N_s-1} \sum_{i=1}^{N_s} \left[T_{i}(t)-T_{\mu}(t)\right]^2
  }.
\end{equation}

\begin{figure*}[p]
\centering
\includegraphics[height=0.837\textheight]{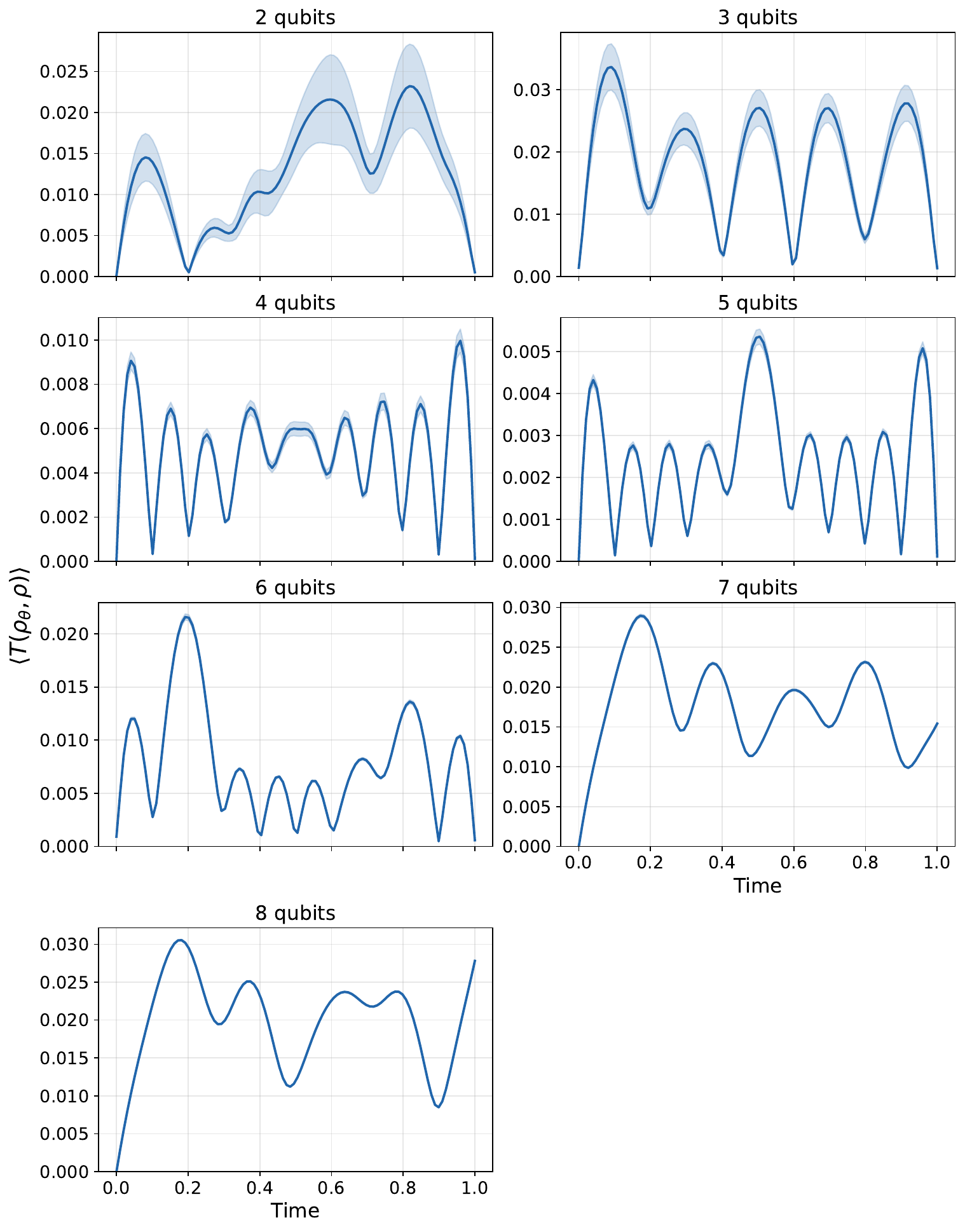}
\caption{
Mean trace distance between states evolved under the exact and predicted unitary operators for system sizes ranging from 2 to 8 qubits. The solid line shows the ensemble average over $N_s=50$ random initial states at each time point, while the shaded region indicates one standard deviation. Trace distances remain of the order of $10^{-2}$ throughout the evolution, demonstrating the accuracy of the learned dynamics.
}
\label{fig:trace-distance-all-N}
\end{figure*}

Figure~\ref{fig:trace-distance-all-N} summarizes the trace-distance statistics for models trained using a dataset with $\Delta t = 0.1$. The obtained values remain consistently of the order of $O(10^{-2})$ across all qubit sizes, indicating a high degree of agreement between the states evolved under the predicted and exact dynamics. This result is particularly significant given that the trace distance is bounded in the interval $[0,1]$ and vanishes if and only if $\rho=\sigma$.

It is also worth noting that all panels in Fig.~\ref{fig:trace-distance-all-N} include a shaded region representing one standard deviation around the ensemble-averaged trace distance. For system sizes of 6 qubits and larger, the variability becomes sufficiently small that the shaded region is barely visible at the scale of the plots. Taken together, the results presented in this section show that the proposed models achieve both high average fidelity and low state-reconstruction error across the considered range of system sizes, highlighting their ability to accurately infer the underlying quantum dynamics.

\subsection{Computational time benchmark}
\label{sec:time-benchmark}
To assess the computational efficiency of the proposed framework, we compare
the inference time of the trained PINN against the geodesic interpolation method
introduced in Ref.~\cite{schilling2024}, which serves as the numerical baseline
throughout this work. Given two known unitary matrices $U_0 = U(t_0)$ and
$U_1 = U(t_1)$ at consecutive time points, the geodesic interpolation constructs
an intermediate unitary at time $t \in [t_0, t_1]$ as
\begin{equation}
U(t) = \left( U_1 U_0^\dagger \right)^\alpha U_0, \quad
\alpha = \frac{t - t_0}{t_1 - t_0},
\end{equation}
where the matrix power is computed via \texttt{fractional\_\allowbreak matrix\_\allowbreak power} from \texttt{scipy.linalg}, which relies on a Schur decomposition and scales as $\mathcal{O}(d^3)$ with the matrix dimension $d = 2^N$.

The benchmark consists of measuring the wall-clock time per query for both methods over 500 independent repetitions per system size, with query times drawn uniformly from $[0, 1]$. For systems of 7- and 8-qubits, the PINN additionally requires a numerical integration step to compute $U(t)$ from the predicted effective Hamiltonian $H_\mathrm{eff}(t)$, using either the Magnus expansion or the Trotterization. Importantly, while $r = 50$ integration steps are used during training, we find that reducing this to $r = 10$ steps at inference time does not lead to a significant loss of accuracy, since the network outputs a smooth effective Hamiltonian for which high temporal resolution is not required at evaluation time. This reduction further decreases the inference cost with respect to the training cost. In this section, all measurements for scenarios up to 5 qubits were performed on an AMD Ryzen 7 5800X, whereas those for cases of 6 qubits and above were performed on a conventional NVIDIA GeForce RTX 3060 GPU.

\begin{figure*}[tp]
\centering
\includegraphics[width=0.99\textwidth]{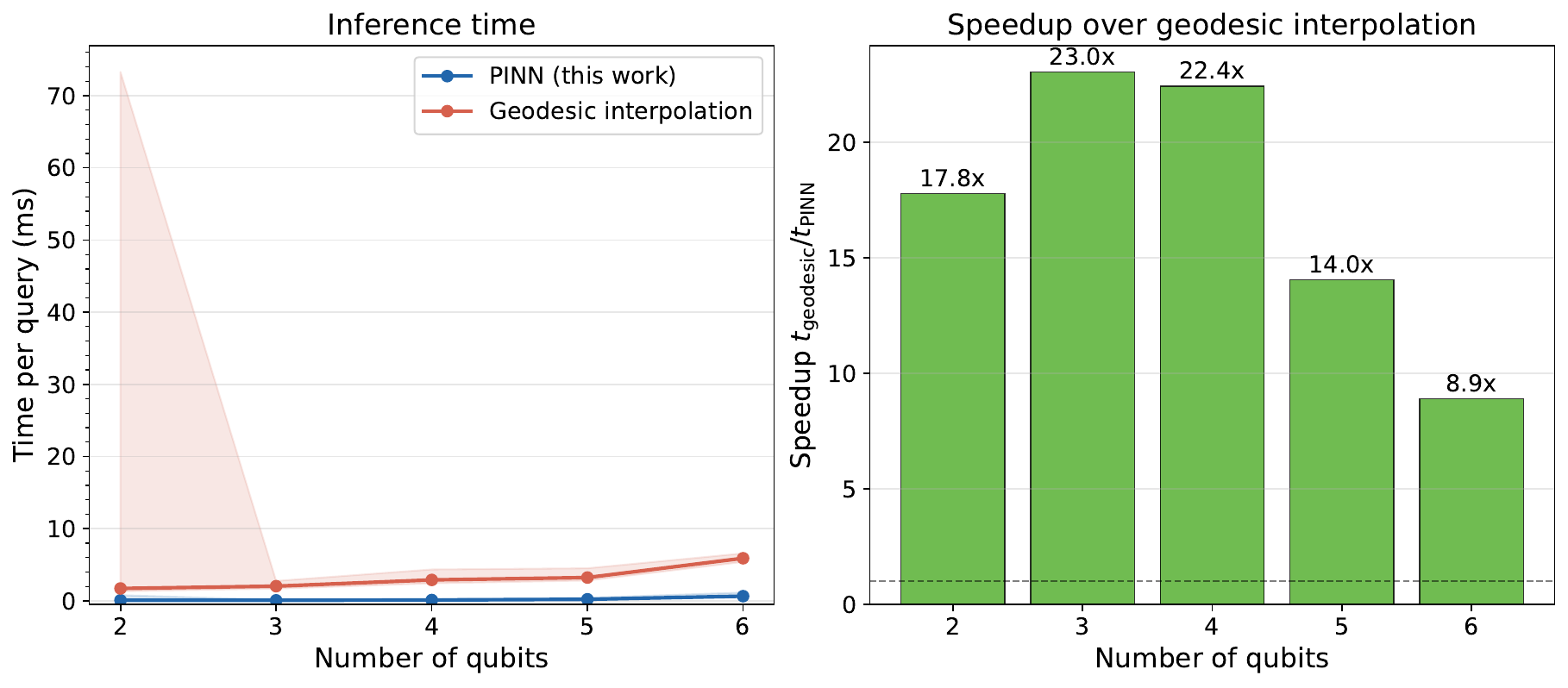}
\caption{
Timing benchmark for systems of 2- to 6-qubits. Left: wall-clock time per query for the PINN and the geodesic interpolation, averaged over 500 repetitions. The shaded region indicates the min/max case. Right: speedup ratio $t_\mathrm{geodesic}/t_\mathrm{PINN}$ for each system size. The dashed line indicates the break-even point at which both methods have equal cost.
}
\label{fig:timing_small}
\end{figure*}

\begin{figure*}[tp]
\centering
\includegraphics[width=0.99\textwidth]{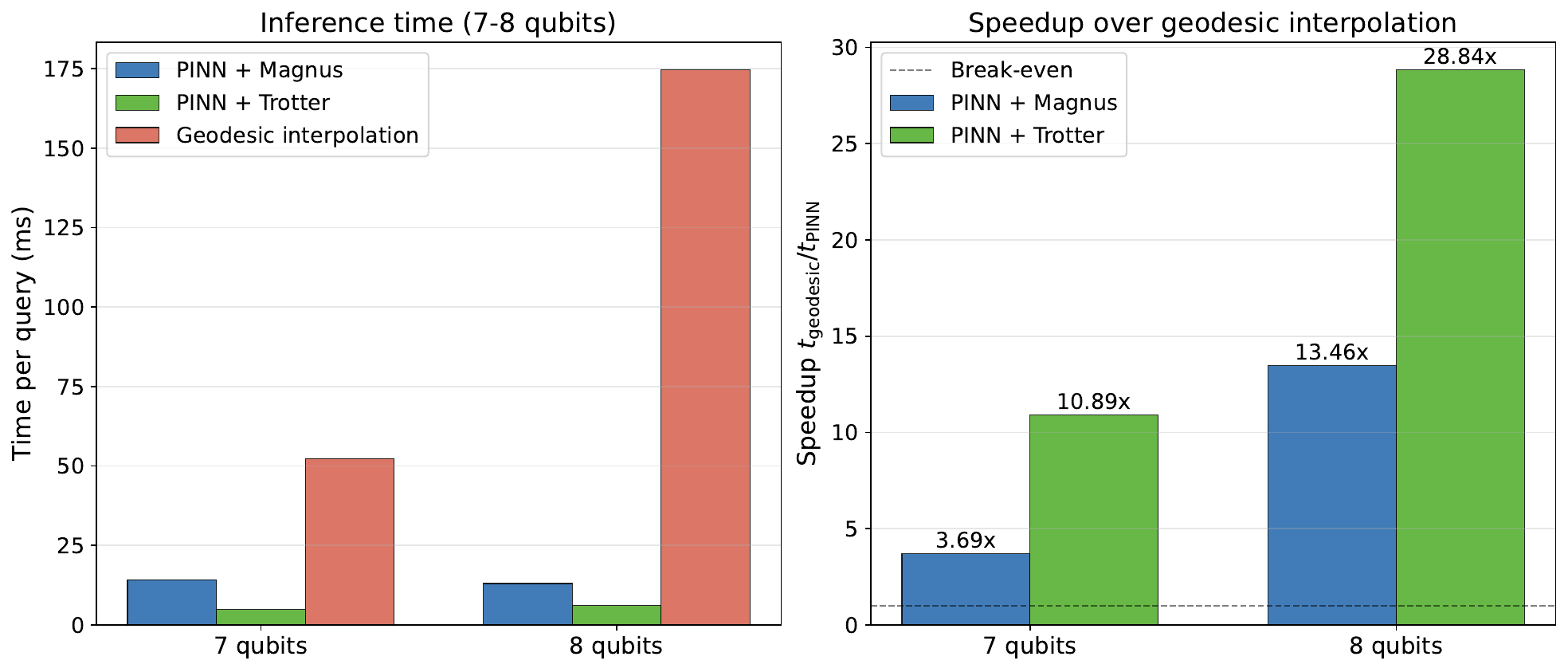}
\caption{
Timing benchmark for systems of 7- and 8-qubits. Left: wall-clock time per query for the PINN variants (Magnus and Trotter) and the geodesic interpolation, averaged over 500 repetitions with $r = 10$ integration steps at inference time. Right: speedup ratio $t_\mathrm{geodesic}/t_\mathrm{PINN}$ for each variant. The dashed line indicates the break-even point.
}
\label{fig:timing_large}
\end{figure*}

The results are summarized in Figs.~\ref{fig:timing_small} and~\ref{fig:timing_large}. For systems of 2- to 6-qubits, the PINN achieves speedups of $8.9\times$--$23.0\times$ over the geodesic interpolation across all system sizes, as shown in Fig.~\ref{fig:timing_small}. The inference time of the PINN remains nearly constant as $N$ increases, while the geodesic interpolation time grows with the matrix dimension, consistent with its $\mathcal{O}(d^3)$ scaling. For systems of 7- and 8-qubits, shown in Fig.~\ref{fig:timing_large}, the Trotter-based variant achieves speedups of $10.9\times$ and $28.8\times$ respectively, while the Magnus-based variant achieves $3.7\times$ and $13.5\times$, the latter being slower due to the additional cost of the nested commutator loop in the second-order correction. In all cases the speedup ratio exceeds unity, confirming that both PINN variants remain faster than the geodesic interpolation. This advantage is expected to grow further for larger systems where the $\mathcal{O}(d^3)$ cost of the geodesic interpolation becomes increasingly prohibitive as $d = 2^N$.

\section{Conclusions and remarks}
\label{sec:conc}
In this work, we introduced a fully connected framework designed to learn and emulate the temporal evolution of many-body quantum systems governed by time-dependent Hamiltonians that are not known a priori. By employing Physics-Informed Neural Networks (PINNs), the model incorporates physical constraints directly into the loss function, enabling it to capture the underlying quantum dynamics. The proposed architecture takes a single scalar input and outputs $2^{2N+1}$ parameters corresponding to the unitary evolution operator. Due to the exponential scaling of the network size with the number of qubits, models were explicitly trained up to 6 qubits. For the 7- and 8-qubit cases, the Hamiltonian was restricted to two physically motivated models: (i) a nearest-neighbor Ising model and (ii) a nearest-neighbor $XYZ$ Heisenberg model, allowing the unitary evolution to be efficiently computed using either a second-order Magnus expansion or a Trotterization. This reduction significantly lowers the number of trainable parameters and renders the problem computationally tractable. The results presented in Section~\ref{sec:results} show that PINNs are capable of accurately interpolating the quantum dynamics encoded in the training data.

To reduce data acquisition costs, we adopted a QPT-based strategy at uniformly spaced time intervals, considering four time-step sizes: $\Delta t = 0.1, 0.25, 0.5$, and $1.0$. As expected, the model trained with the smallest time step achieved the highest fidelity. Remarkably, models trained with $\Delta t = 1.0$ achieved comparable fidelities, suggesting that sparse temporal sampling may already be sufficient to reproduce the system dynamics. The optimal choice of $\Delta t$ is expected to depend on the Hamiltonian's time dependence: systems with strong nonlinearities or high-frequency components may require finer discretization. Generalization over extended temporal domains remains challenging for neural-network models~\cite{zhou2025, li2025}, so some performance degradation at larger time intervals is expected, though it can be mitigated by increasing the number of QPT samples. Two further reductions are examined in \ref{sec:app4}: removing the unitary supervision term, so that training requires state tomography rather than full process tomography, and lowering the number of training samples by one to two orders of magnitude with respect to the value used here.

In addition, a statistical analysis based on multiple independent runs with different Hamiltonians was performed to assess the robustness of the proposed approach. The results show that the models consistently achieve mean fidelities close to $0.99$, with very low standard deviations. No specific time regions were identified in which the interpolation systematically fails or exhibits consistently low fidelity, indicating that the model is able to reliably learn the quantum dynamics given sufficient training data.

A related problem was analyzed in Ref.~\cite{schilling2024}, where a mathematical framework for interpolating unitary matrices using numerical techniques was proposed, yielding promising results. Compared to that approach, the method introduced here offers greater flexibility and can be readily extended to scenarios involving larger system sizes, different Hamiltonians, additional physical constraints, or nonlinear parameter dependencies. Moreover, since the numerical method relies on explicit fractional matrix power, which becomes computationally expensive for large systems, the framework presented in this work provides a potentially more efficient alternative in applications requiring frequent interpolation of unitary evolution operators, as shown in Section~\ref{sec:time-benchmark}.

Despite its promising performance, the proposed framework presents several limitations that define its range of applicability. The model must be retrained for each Hamiltonian and for each system size, since the network architecture depends on the number of qubits. Consequently, the method is intended for studying the dynamics of a specific physical system rather than generalizing across different interactions or system sizes. However, these requirements are not as restrictive as they may appear. In most physical scenarios, both the number of qubits and the functional form of $H(t)$ are fixed by the experimental platform, while only the quantum state evolves in time. A model tailored to a given Hamiltonian and system size can therefore be highly practical for repeatedly evaluating the dynamics of the same system. The method may also struggle in regimes with highly oscillatory or rapidly varying dynamics, where larger architectures would likely be required. Furthermore, while inference is faster than conventional numerical solvers, this advantage is most significant when the dynamics must be evaluated repeatedly, for instance in dynamical quantum state tomography or parameter-estimation tasks. Finally, for the 7- and 8-qubit cases, the learned effective Hamiltonian does not coincide with the target Hamiltonian, indicating that the method captures the underlying dynamics without performing Hamiltonian reconstruction. Extending the framework to explicitly learn the true Hamiltonian constitutes a promising direction for future research.

Overall, the results of this study highlight the potential of deep learning for modeling quantum dynamics, particularly in regimes where data are sparse or discretized. Unlike approaches that explicitly reconstruct the Hamiltonian from equations of motion~\cite{greydanus2019hamiltonian, wang2023machine}, the proposed framework is trained to reproduce quantum dynamics directly from state evolution data. In this sense, the present work is conceptually closer to operator-learning approaches, where the objective is to approximate the propagator itself rather than identify the underlying physical model~\cite{shah2025machine}. Although an effective Hamiltonian can be inferred from the learned dynamics, the method is not designed to recover the exact Hamiltonian parameters; it aims instead to accurately reproduce the resulting time evolution, even when only partial structural information about the Hamiltonian is available. Future work may explore architectures capable of handling multiple system sizes within a unified model and investigate systems with strong time dependence or chaotic behavior to further assess robustness and generalization. Building on the differentiability of the model output, a natural line of continuity is to enrich the framework with additional physical constraints: incorporating differential constraints derived from the Schr\"{o}dinger equation may provide a pathway toward Hamiltonian reconstruction while preserving the flexibility of the current data-driven approach, and the extension to open quantum dynamics offers a complementary setting in which further structure can be imposed on the learned evolution. Additionally, integrating this methodology with continuous-measurement protocols could further reduce experimental overhead by exploiting the model's ability to generate reliable predictions at arbitrary time points.

\section*{Acknowledgments}
The authors gratefully acknowledge the financial support provided by ANID (Agencia Nacional de Investigación y Desarrollo) through the doctoral scholarship No. 2022-21221096, the research project No. 2023-3230427, FONDECYT 1231940, and the ANID Millennium Science Initiative Program under grant No. ICN17\_012. This support has been essential to the development and completion of this research.

\appendixstart
\label{sec:app}

\section{Numerical integration methods}
\label{sec:app-integration}

This appendix describes the two numerical methods used to compute the time-evolution operator $U(t)$ from the effective Hamiltonian $H_\mathrm{eff}(t)$ predicted by the neural network for systems of 7- and 8-qubits. Both methods guarantee exact unitarity of the output by construction, rendering the explicit unitarity penalty in the loss function unnecessary for these cases.

\subsection*{Second-order Magnus expansion}
The Magnus expansion expresses the time-evolution operator as
\begin{equation}
  \label{eq:magnus-exp}
  U(t) = \exp\!\left( \sum_{k=1}^{\infty} \Omega_k(t) \right),
\end{equation}
where, truncated at second order, the exponent terms are
\begin{eqnarray}
  \Omega_1(t) &=& -\frac{i}{\hbar} \int_{t_0}^{t} H(\tau_1) \, \mathrm{d}\tau_1, \\
  \Omega_2(t) &=& -\frac{1}{2\hbar^2} \int_{t_0}^{t} \mathrm{d}\tau_1
  \int_{t_0}^{\tau_1} \mathrm{d}\tau_2 \left[ H(\tau_1), H(\tau_2) \right].
\end{eqnarray}
This truncation is valid when the Magnus series converges, which is
guaranteed by the sufficient condition~\cite{blanes2009}
\begin{equation}
\int_0^T \|H(t)\|_{\mathrm{s}} \, dt < \pi,
\end{equation}
where $\|H(t)\|_{\mathrm{s}}$ denotes the spectral norm of the Hamiltonian at time $t$.

Given a target time $t$ and a number of integration steps $r$, a uniform
time grid $\{t_k\}_{k=0}^{r-1}$ is constructed over $[0, t]$ with step
size $\Delta t = t/r$. The neural network $f_\theta$ is evaluated at each
grid point to obtain the sequence of Hamiltonians $\{H(t_k)\}_{k=0}^{r-1}$.
The Magnus operator $\Omega(t) \approx \Omega_1(t) + \Omega_2(t)$ is then
approximated as
\begin{eqnarray}
    \Omega_1(t) &=& \sum_{k=0}^{r-1} H(t_k)\, \Delta t, \\
    \Omega_2(t) &=& -\frac{i}{2} \sum_{i=0}^{r-1} \sum_{j=0}^{i-1}
    \left[ H(t_i),\, H(t_j) \right] \Delta t^2,
\end{eqnarray}
and the time-evolution operator is recovered as
\begin{equation}
    U(t) = \exp\!\left(-i\,\Omega(t)\right).
\end{equation}
The full procedure is summarized in Algorithm~\ref{alg:magnus}.

\begin{algorithm}[H]
\caption{Second-order Magnus expansion}
\label{alg:magnus}
\begin{algorithmic}[1]
\REQUIRE Model $f_\theta$, target time $t$, operator basis $\{\sigma_\mu\}$,
         number of steps $r$
\ENSURE Unitary operator $U(t) \in \mathrm{SU}(d)$
\STATE $\Delta t \leftarrow t / r$
\FOR{$k = 0$ \TO $r-1$}
    \STATE $t_k \leftarrow k\,\Delta t$
    \STATE $\boldsymbol{c}_k \leftarrow f_\theta(t_k)$ \COMMENT{model forward pass}
    \STATE $H(t_k) \leftarrow \sum_\mu c_k^\mu\,\sigma_\mu$ \COMMENT{reconstruct Hamiltonian}
\ENDFOR
\STATE $\Omega_1 \leftarrow \sum_{k=0}^{r-1} H(t_k)\,\Delta t$
\STATE $\Omega_2 \leftarrow \mathbf{0}$
\FOR{$i = 0$ \TO $r-1$}
    \FOR{$j = 0$ \TO $i-1$}
        \STATE $\Omega_2 \leftarrow \Omega_2
            - \frac{i}{2}\left[H(t_i),\,H(t_j)\right]\Delta t^2$
    \ENDFOR
\ENDFOR
\STATE $U(t) \leftarrow \exp\!\left(-i\,(\Omega_1 + \Omega_2)\right)$
\RETURN $U(t)$
\end{algorithmic}
\end{algorithm}

\subsection*{Trotterization}
The Trotterization approximates the time-ordered exponential \cite{trotter1959} by discretizing the time interval into $r$ subintervals of equal length $\Delta t = t/r$ and freezing the Hamiltonian within each slice,
\begin{equation}
  \label{eq:trotter}
  U(t) \approx \prod_{k=r-1}^{0} \exp\!\left(-i\,H(t_k)\,\Delta t\right),
\end{equation}
where the product is taken in chronological order (right to left). To
improve accuracy over simple left-endpoint sampling, the Hamiltonian is
evaluated at the midpoint of each subinterval,
\begin{equation}
    t_k = \left(k + \frac{1}{2}\right)\frac{t}{r},
    \quad k = 0, 1, \ldots, r-1.
\end{equation}
Since each factor $\exp(-i H(t_k)\Delta t)$ is exactly unitary, their
product is also exactly unitary regardless of the step size, which is the
key practical advantage of this method. The full procedure is summarized
in Algorithm~\ref{alg:trotter}.

\begin{algorithm}[H]
\caption{First-order Lie--Trotter decomposition}
\label{alg:trotter}
\begin{algorithmic}[1]
\REQUIRE Model $f_\theta$, target time $t$, operator basis $\{\sigma_\mu\}$,
         number of steps $r$
\ENSURE Unitary operator $U(t) \in \mathrm{SU}(d)$
\STATE $\Delta t \leftarrow t / r$
\STATE $U \leftarrow \mathbf{I}_d$
\FOR{$k = 0$ \TO $r-1$}
    \STATE $t_k \leftarrow \left(k + \frac{1}{2}\right)\Delta t$
           \COMMENT{midpoint sampling}
    \STATE $\boldsymbol{c}_k \leftarrow f_\theta(t_k)$
           \COMMENT{model forward pass}
    \STATE $H(t_k) \leftarrow \sum_\mu c_k^\mu\,\sigma_\mu$
    \STATE $U_k \leftarrow \exp\!\left(-i\,H(t_k)\,\Delta t\right)$
    \STATE $U \leftarrow U_k\,U$
           \COMMENT{left-multiply to preserve time ordering}
\ENDFOR
\RETURN $U$
\end{algorithmic}
\end{algorithm}

\noindent
In both methods, $r = 50$ integration steps are used during training and $r = 10$ steps at inference time, since the neural network outputs a smooth effective Hamiltonian for which high temporal resolution is not required at evaluation time.

\section{Model architecture}
\label{sec:app1}

\subsection*{Cases with 2 to 6 qubits}
The model's internal data processing involves several fully connected layers. Initially, the input layer takes a scalar input corresponding to time. This input is processed through fully connected layers that progressively increase the dimensionality of the latent representation. After each hidden layer, a hyperbolic tangent activation function is applied. The resulting features are then reshaped into tensors of dimension $(2, 2^N, 2^N)$, where the first index is used to construct the real and imaginary parts of the resulting complex-valued matrix. A detailed description of each architecture is presented in Table~\ref{tab:model-architecture}.

\begin{table*}[t]
\centering
\renewcommand{\arraystretch}{1.2}

\begin{minipage}[t]{0.48\textwidth}
\centering
\textbf{\textit{model2}} (2 qubits)\\[0.5ex]
\begin{tabular}{c|c|c}
\hline
\textbf{Layer} & \textbf{Input} & \textbf{Output} \\
\hline \hline
Linear 1 & 1 & 64 \\
Tanh & - & - \\
Linear 2 & 64 & 128 \\
Tanh & - & - \\
Linear 3 & 128 & 32 \\
Reshape & 32 & $(2,4,4)$ \\
\hline
\end{tabular}
\end{minipage}
\hfill
\begin{minipage}[t]{0.48\textwidth}
\centering
\textbf{\textit{model3}} (3 qubits)\\[0.5ex]
\begin{tabular}{c|c|c}
\hline
\textbf{Layer} & \textbf{Input} & \textbf{Output} \\
\hline \hline
Linear 1 & 1 & 256 \\
Tanh & - & - \\
Linear 2 & 256 & 512 \\
Tanh & - & - \\
Linear 3 & 512 & 128 \\
Reshape & 128 & $(2,8,8)$ \\
\hline
\end{tabular}
\end{minipage}

\medskip

\begin{minipage}[t]{0.48\textwidth}
\centering
\textbf{\textit{model4}} (4 qubits)\\[0.5ex]
\begin{tabular}{c|c|c}
\hline
\textbf{Layer} & \textbf{Input} & \textbf{Output} \\
\hline \hline
Linear 1 & 1 & 512 \\
Tanh & - & - \\
Linear 2 & 512 & 1024 \\
Tanh & - & - \\
Linear 3 & 1024 & 2048 \\
Tanh & - & - \\
Linear 4 & 2048 & 512 \\
Reshape & 512 & $(2,16,16)$ \\
\hline
\end{tabular}
\end{minipage}
\hfill
\begin{minipage}[t]{0.48\textwidth}
\centering
\textbf{\textit{model5}} (5 qubits)\\[0.5ex]
\begin{tabular}{c|c|c}
\hline
\textbf{Layer} & \textbf{Input} & \textbf{Output} \\
\hline \hline
Linear 1 & 1 & 512 \\
Tanh & - & - \\
Linear 2 & 512 & 2048 \\
Tanh & - & - \\
Linear 3 & 2048 & 2048 \\
Reshape & 2048 & $(2,32,32)$ \\
\hline
\end{tabular}
\end{minipage}

\medskip

\begin{minipage}[t]{0.48\textwidth}
\centering
\textbf{\textit{model6}} (6 qubits)\\[0.5ex]
\begin{tabular}{c|c|c}
\hline
\textbf{Layer} & \textbf{Input} & \textbf{Output} \\
\hline \hline
Linear 1 & 1 & 1024 \\
Tanh & - & - \\
Linear 2 & 1024 & 4096 \\
Tanh & - & - \\
Linear 3 & 4096 & 8192 \\
Reshape & 8192 & $(2,64,64)$ \\
\hline
\end{tabular}
\end{minipage}

\caption{
Architectures of \textit{model2} through \textit{model6}, each representing the generator of a $2^N \times 2^N$ complex-valued unitary matrix from a scalar temporal input $t$.
}
\label{tab:model-architecture}
\end{table*}

\subsection*{Cases with 7 and 8 qubits}

For the models handling 7 and 8 qubits, the architecture follows a similar structure to the previous models but differs significantly in the output layer. Instead of generating a complex-valued matrix directly, the output layer produces time-dependent coefficients associated with the Pauli basis representation of the Hamiltonian $H(t)$. A detailed description of these architectures is presented in Table~\ref{tab:model-architecture2}.

\begin{table*}[t]
\centering
\renewcommand{\arraystretch}{1.2}

\begin{minipage}[t]{0.48\textwidth}
\centering
\textbf{\textit{model7}} (7 qubits)\\[0.5ex]
\begin{tabular}{c|c|c}
\hline
\textbf{Layer} & \textbf{Input} & \textbf{Output} \\
\hline \hline
Linear 1 & 1 & 50 \\
Tanh & - & - \\
Linear 2 & 50 & 13 \\
\hline
\end{tabular}
\end{minipage}
\hfill
\begin{minipage}[t]{0.48\textwidth}
\centering
\textbf{\textit{model8}} (8 qubits)\\[0.5ex]
\begin{tabular}{c|c|c}
\hline
\textbf{Layer} & \textbf{Input} & \textbf{Output} \\
\hline \hline
Linear 1 & 1 & 50 \\
Tanh & - & - \\
Linear 2 & 50 & 15 \\
\hline
\end{tabular}
\end{minipage}

\caption{
Architectures of \textit{model7} and \textit{model8}, designed to generate real-valued coefficients associated with the Pauli basis representation of the Hamiltonian from a scalar temporal input $t$.
}
\label{tab:model-architecture2}
\end{table*}


\section{Unitarity imposition term}
\label{sec:app2}
In Section~\ref{sec:loss-function}, we introduced the loss function employed during training together with a brief description of each contribution and its physical interpretation. The first two terms naturally incorporate the available knowledge of the system through supervised data constraints. In contrast, the third term introduces an additional soft constraint associated with the unitarity properties of the time-evolution operator, which may not always play a dominant role during the learning process.

\begin{figure*}[tp]
\centering
\includegraphics[width=0.99\textwidth]{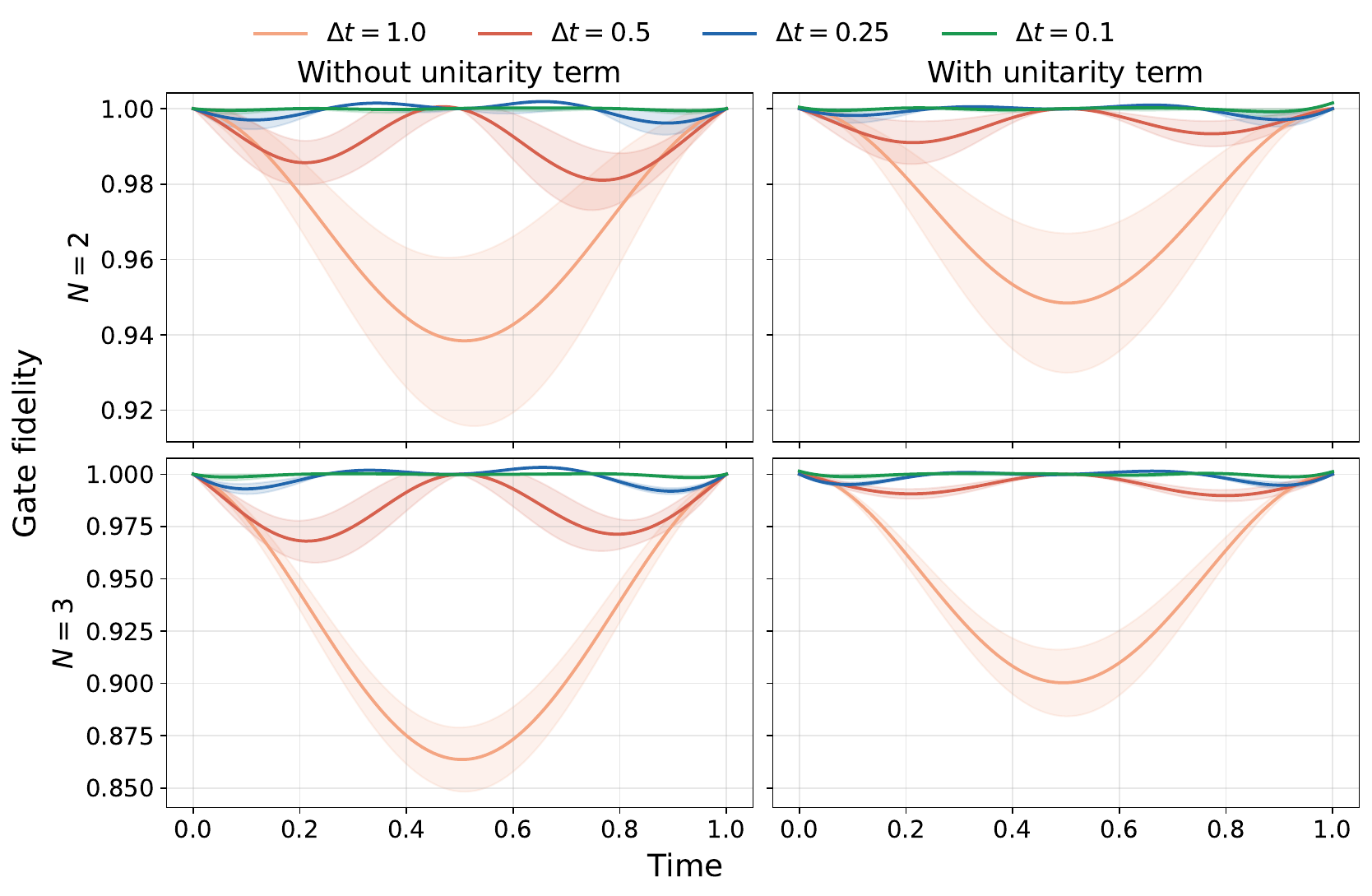}
\caption{
Statistical analysis of the effect of the unitarity imposition term in the loss function over 25 independent runs, for the $N=2$ and $N=3$ qubit cases. In each row, the left panel shows the results obtained without the unitarity term, while the right panel includes the additional soft constraint associated with the unitary behavior of the time-evolution operator. Each curve represents the mean gate fidelity as a function of time for different temporal resolutions $\Delta t = 0.1, 0.25, 0.5,$ and $1.0$, while the shaded regions indicate the corresponding standard deviation across independent realizations with randomly generated Hamiltonian parameters.
}
\label{fig:unitarity_analysis_1}
\end{figure*}

\begin{figure*}[tp]
\centering
\includegraphics[width=0.99\textwidth]{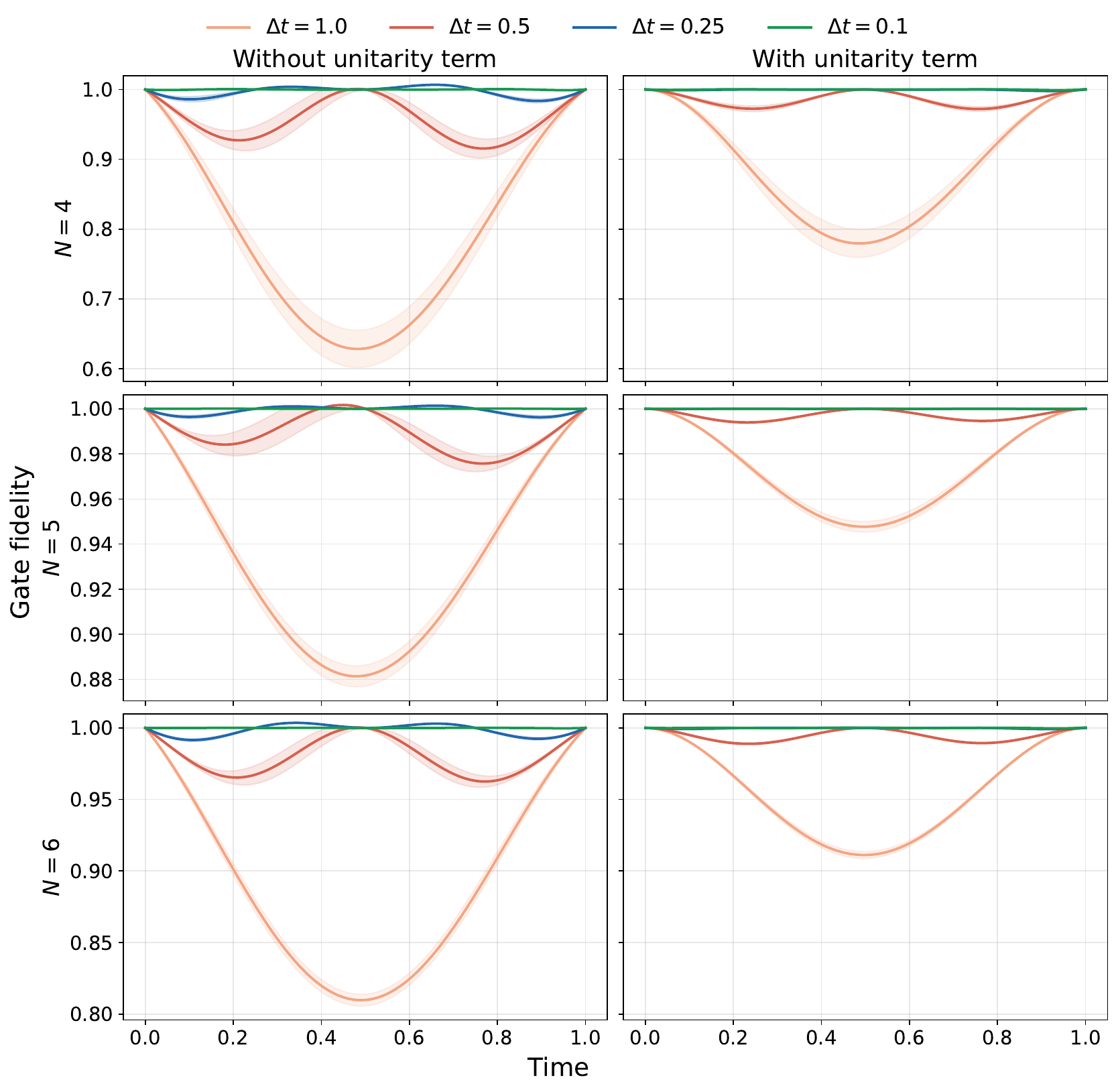}
\caption{
Same as Fig.~\ref{fig:unitarity_analysis_1}, for the $N=4$, $N=5$, and $N=6$ qubit cases.
}
\label{fig:unitarity_analysis_2}
\end{figure*}

Figures~\ref{fig:unitarity_analysis_1}--\ref{fig:unitarity_analysis_2} present a statistical analysis of the impact of including or excluding the unitarity penalty term during training. The analysis was carried out through 25 independent training runs per system size, each considering a distinct realization of the fully non-zero Pauli Hamiltonian, with parameters sampled as described in Section~\ref{sec:n-qubit-systems}. As in Section~\ref{sec:results}, four values of the time grid were considered, $\Delta t \in \{0.1, 0.25, 0.5, 1.0\}$.

The figures compare the mean gate fidelity and its associated standard deviation across multiple independent realizations. The results show that the inclusion of the unitarity term becomes increasingly relevant as the amount of available training data decreases. In particular, for cases with sufficiently dense temporal sampling, such as $\Delta t = 0.1$, both approaches achieve fidelities close to unity over the entire time interval, indicating that the additional unitarity constraint does not provide improvements for the model. For $\Delta t = 0.25$, a small but still noticeable improvement can already be observed when the unitarity term is included.

However, for cases with more limited information, namely $\Delta t = 0.5$ and $\Delta t = 1.0$, the inclusion of the unitarity term significantly improves the overall performance of the model. In these regimes, the additional constraint not only increases the mean fidelity but also reduces the variance across independent realizations, leading to more stable and reliable predictions. This behavior is physically reasonable since, under reduced data availability, one of the few pieces of prior information that remains known with certainty is the unitary nature of the time-evolution operator. Consequently, these results highlight the importance of incorporating physically motivated constraints into the loss function, particularly in low-data regimes.

\section{Gate fidelity between unitary operators}
\label{sec:app3}

There exist different useful metrics to compare unitary matrices. In particular, we choose the gate fidelity because it is a natural metric that can be derived from the quantum state fidelity between the Choi matrices of the quantum maps associated with the unitary operators being compared. To prove this statement, it is convenient to adopt the framework of quantum channels: instead of comparing the unitary matrices directly, we define a quantum map that acts via each unitary operator. A quantum operation can be represented as a linear completely positive trace-preserving map $\mathcal{E}$ acting on density operators according to
\begin{equation}
\rho' = \mathcal{E}(\rho).
\end{equation}

This representation is particularly useful for comparing quantum channels, and a natural tool for this purpose is provided by the Choi--Jamio\l{}kowski isomorphism \cite{Watrous2018}, which establishes a one-to-one correspondence between completely positive maps and positive semidefinite operators acting on an enlarged Hilbert space.

Consider a unitary operator $U$ acting on a $d$-dimensional Hilbert space. The associated unitary quantum channel $\mathcal{U}$ is defined as
\begin{equation}
\mathcal{U}(\rho) = U \rho U^\dagger.
\end{equation}
The corresponding Choi matrix is given by
\begin{equation}
J_U = (\mathcal{U} \otimes \mathbb{I})(|\Phi\rangle\langle\Phi|),
\end{equation}
where
\begin{equation}
|\Phi\rangle = \frac{1}{\sqrt{d}} \sum_{i=1}^{d} |i\rangle \otimes |i\rangle
\end{equation}
is a maximally entangled reference state.

Since the Choi--Jamio\l{}kowski isomorphism maps quantum operations to density operators, the fidelity between two quantum channels can be defined as the fidelity between their respective Choi matrices. Consider two unitary operators $U$ and $V$, with associated Choi matrices
\begin{eqnarray}
J_U &=& (\mathcal{U} \otimes \mathbb{I})(|\Phi\rangle \langle\Phi|), \\
J_V &=& (\mathcal{V} \otimes \mathbb{I})(|\Phi\rangle \langle\Phi|).
\end{eqnarray}
For unitary channels, the Choi matrices correspond to pure states $|\Phi_U\rangle$ and $|\Phi_V\rangle$, so the fidelity between them reduces to
\begin{equation}
F(J_U, J_V) = |\langle \Phi_U | \Phi_V \rangle|^2.
\end{equation}
Using the explicit form of the Choi states, this expression simplifies to
\begin{equation}
F(J_U, J_V)
=
\left|
\langle \Phi |
(\mathbb{I} \otimes U^\dagger V)
| \Phi \rangle
\right|^2
=
\frac{1}{d^2}
\left|
\mathrm{Tr}(U^\dagger V)
\right|^2,
\end{equation}
which is conventionally referred to as the \textit{gate fidelity} \cite{Watrous2018,cabrera2011}. In particular, when $U = V$, one obtains $F(J_U, J_V) = 1$, corresponding to perfect agreement between both unitary operations. Therefore, the gate fidelity provides a consistent and operationally meaningful measure for quantifying the similarity between two unitary quantum evolutions, and it is the metric adopted throughout this work.

\section{Training with partial information}
\label{sec:app4}

The loss function in Eq.~\eqref{eq:loss} contains two supervised contributions. The first compares the predicted and target evolved states and therefore requires only pairs $(\rho_0^i,\rho_{\tau_i}^i)$, which can in principle be obtained through state tomography at the sampled times. The second directly compares the predicted evolution operator with the target unitary $U_{\tau_i}$. This term therefore requires knowledge of the evolution operator at the sampled times, although not throughout the full time interval. While such information is available in our numerical experiments and represents the most favorable supervision scenario considered in the main text, it need not be available when characterizing an experimental system with an unknown Hamiltonian. We therefore assess whether direct unitary supervision is necessary for accurate interpolation.

To this end, we compare two training scenarios that differ only in the supervised information entering the loss. In the first, denoted \textit{states only}, the unitary term is removed and no target evolution operator is provided to the network during training. In the second, denoted \textit{states $+$ unitaries}, both supervised terms are retained, as in the main text. The Hamiltonian realization, model architecture, optimizer, training budget, temporal grids, and evaluation protocol are otherwise identical between the two scenarios.

The comparison is performed using the nearest-neighbor Ising family over system sizes for which both modeling strategies can be evaluated. For the direct ansatz, the network predicts $U(t)$ for $N=2$ to $6$, whereas for the effective-Hamiltonian ansatz it predicts $H_{\mathrm{eff}}(t)$ for $N=2$ to $8$, from which the evolution operator is obtained by Trotterization. The time-dependent coefficients follow Eq.~\eqref{eq:coeff-form}, with $\omega\in[0,6\pi)$ and an amplitude scaling that decreases with system size, from $2.08$ for $N=2$ to $1.34$ for $N=8$. These parameters produce a more strongly coupled dynamical regime than that considered in the main text, reducing the possibility that high fidelities arise simply from weak temporal variation of the evolution operator. Target unitaries are generated using 50 Trotter slices. Training is performed for 400 epochs using the scheduler described in Sec.~\ref{sec:test-train}, with $N_{\rho}=1000$ samples for the direct ansatz and $N_{\rho}=100$ for the effective-Hamiltonian ansatz. The reported results are averaged over 20 and 10 independent Hamiltonian realizations, respectively.

\begin{figure*}[tp]
\centering
\includegraphics[width=0.99\textwidth]{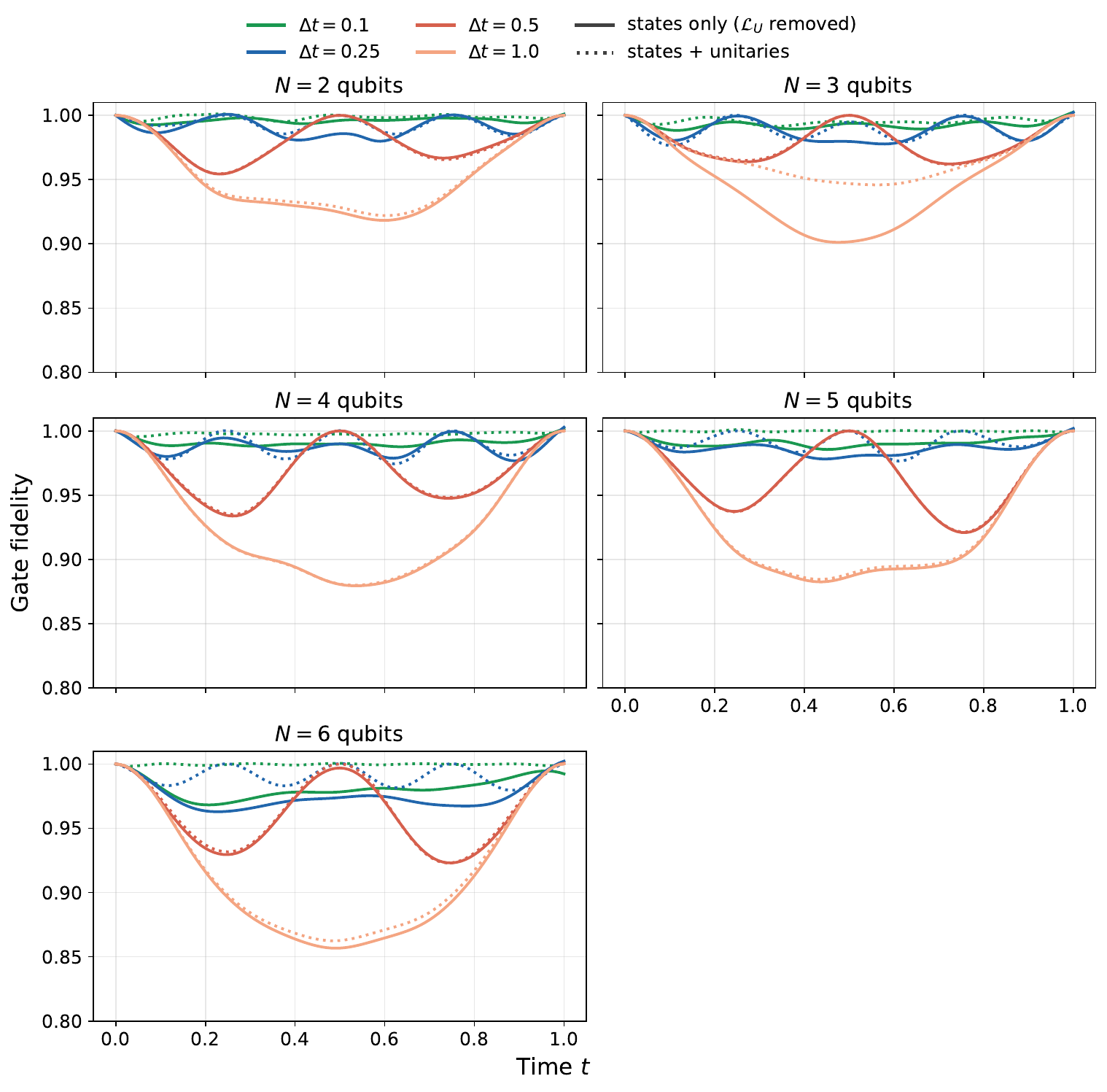}
\caption{
Mean gate fidelity for the direct ansatz, which predicts the evolution operator $U(t)$, for nearest-neighbor Ising systems ranging from 2 to 6 qubits. Solid lines correspond to training using state data only, whereas dotted lines correspond to training using both state and unitary data. Each curve is averaged over 20 independent Hamiltonian realizations.
}
\label{fig:loss-ablation-U}
\end{figure*}

\begin{figure*}[p]
\centering
\includegraphics[height=0.825\textheight]{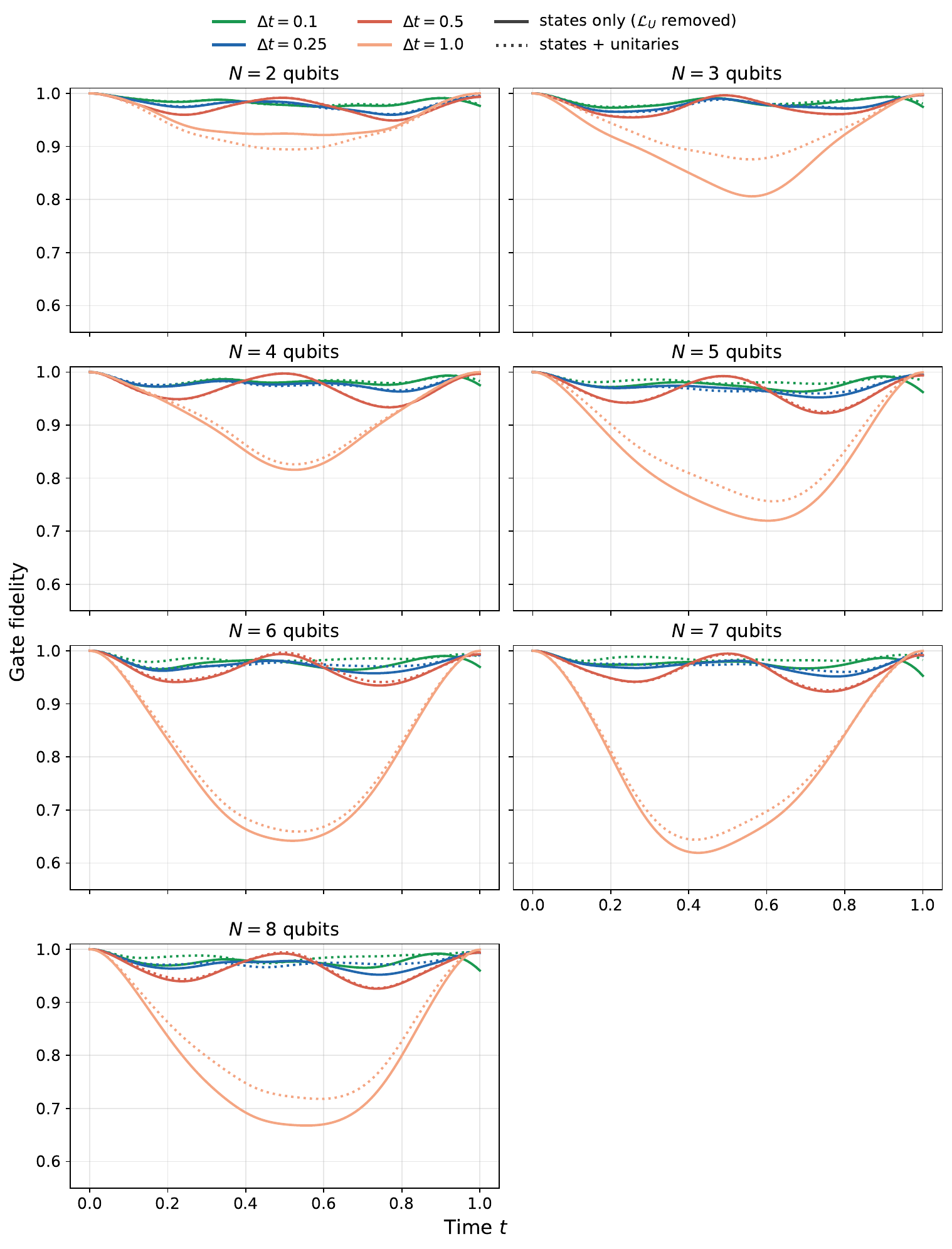}
\caption{
Mean gate fidelity for the effective-Hamiltonian ansatz, in which the network predicts the coefficients of $H_{\mathrm{eff}}(t)$ and the corresponding evolution operator is obtained by Trotterization. Results are shown for nearest-neighbor Ising systems ranging from 2 to 8 qubits. Solid lines correspond to training using state data only, whereas dotted lines correspond to training using both state and unitary data. Each curve is averaged over 10 independent Hamiltonian realizations.
}
\label{fig:loss-ablation-Heff}
\end{figure*}

Figures~\ref{fig:loss-ablation-U} and~\ref{fig:loss-ablation-Heff} show the resulting mean gate fidelity over the full time domain. For $\Delta t=0.1$, $0.25$, and $0.5$, the two training scenarios yield nearly indistinguishable fidelity curves across all system sizes and for both ansatzes. The corresponding time-averaged differences are at most $0.018$ for the direct ansatz and $0.008$ for the effective-Hamiltonian ansatz, with no systematic advantage for either training scenario. The dispersion across Hamiltonian realizations remains below $0.025$ for all these configurations. It is slightly larger when only state data are provided, consistent with the reduced amount of supervision available during training, and is omitted from the figures for clarity.

A noticeable difference emerges only for $\Delta t=1.0$, corresponding to the sparsest temporal sampling and therefore to the least constrained interpolation problem. In this regime, including the target unitaries improves the mean fidelity by up to $0.03$. This behavior is consistent with the role of the unitarity constraint discussed in \ref{sec:app2}, which likewise becomes more relevant when the available training information is strongly reduced. Thus, direct unitary supervision provides additional regularization in the low-data regime but has little effect once the temporal sampling is sufficiently informative.

These experiments also provide insight into the amount of training data required. The results in the main text were obtained using $N_{\rho}=11000$, whereas the present analysis uses only $1000$ samples for the direct ansatz and $100$ for the effective-Hamiltonian ansatz. Despite this reduction of one to two orders of magnitude, the mean fidelities remain above $0.96$ for $\Delta t\leq0.5$, even in the more strongly coupled regime considered here. This comparison is not a controlled data-efficiency study because the coupling strengths and number of training epochs also differ from those of the main text. Nevertheless, it indicates that the original dataset size is substantially larger than what is required to obtain high interpolation fidelity under the conditions tested here. Determining the minimum amount of characterization data required to reach a prescribed fidelity constitutes an experimentally relevant optimization problem that we leave for future work.

Overall, these results demonstrate that direct supervision with target evolution operators is not required for accurate interpolation when the temporal sampling provides sufficient information. Comparable fidelities can be obtained using only pairs of initial and evolved states, without presenting any target unitary to the network during training. This is particularly relevant to the experimental setting motivating our approach, in which the underlying Hamiltonian is unknown and the available information may consist only of state measurements at discrete times. Direct unitary supervision can nevertheless be incorporated when process-characterization data are available and becomes beneficial in the sparsely sampled regime. The interpolation framework itself, however, does not depend on such information.



\onecolumn
\bibliographystyle{unsrt}
\bibliography{bio.bib}

\end{document}